\documentclass[reqno,10pt,a4paper,dvips]{amsart}

\usepackage{amssymb,mathptmx,cite,psfrag,eucal,array,setspace,geometry,enumitem}
\usepackage[dvips]{graphicx}
\usepackage{tikz}

\numberwithin{equation}{section}

\newcolumntype{C}{>{$}c<{$}} 
\allowdisplaybreaks

\newcommand{\alg}[1]{\mathfrak{#1}}

\newcommand{\func}[2]{#1 \left( #2 \right)}
\newcommand{\tfunc}[2]{#1 \bigl( #2 \bigr)}

\newcommand{\brac}[1]{\left( #1 \right)}
\newcommand{\sqbrac}[1]{\left[ #1 \right]}
\newcommand{\set}[1]{\left\{ #1 \right\}}

\newcommand{\abs}[1]{\left| #1 \right|}

\newcommand{\ZZ}{\mathbb{Z}}

\newcommand{\RR}{\mathbb{R}}
\newcommand{\CC}{\mathbb{C}}

\newcommand{\ii}{\mathfrak{i}}

\newcommand{\wun}{\mathbf{1}}

\newcommand{\killing}[2]{\kappa \bigl( #1 , #2 \bigr)}

\newcommand{\affine}[1]{\widehat{#1}}

\newcommand{\comm}[2]{\bigl[ #1 , #2 \bigr]}

\newcommand{\ket}[1]{\bigl\lvert #1 \bigr\rangle}
\newcommand{\braket}[2]{\bigl\langle #1 \bigr\rvert \bigl. #2 \bigr\rangle}
\newcommand{\bracket}[3]{\bigl\langle #1 \bigr\rvert #2 \bigl\lvert #3 \bigr\rangle} 

\newcommand{\normord}[1]{{} : #1 : {}} 

\newcommand{\AffIrrMod}[1]{\affine{\mathcal{L}}_{#1}}

\newcommand{\AffOthMod}[1]{\affine{\mathcal{E}}_{#1}}
\newcommand{\AffStagMod}[1]{\affine{\mathcal{S}}_{#1}}

\newcommand{\SLA}[2]{\alg{#1} \left( #2 \right)}
\newcommand{\AKMA}[2]{\affine{\alg{#1}} \left( #2 \right)}

\newcommand{\fuse}{\mathbin{\times_{\! f}}}

\newcommand{\dses}[3]{0 \longrightarrow #1 \longrightarrow #2 \longrightarrow #3 \longrightarrow 0}

\newcommand{\eqnref}[1]{Equation~\eqref{#1}}
\newcommand{\eqnDref}[2]{Equations~\eqref{#1} and \eqref{#2}}

\newcommand{\secref}[1]{Section~\ref{#1}}
\newcommand{\secDref}[2]{Sections~\ref{#1} and \ref{#2}}

\newcommand{\figref}[1]{Figure~\ref{#1}}
\newcommand{\figDref}[2]{Figures~\ref{#1} and \ref{#2}}

\newcommand{\cft}{conformal field theory}
\newcommand{\cfts}{conformal field theories}
\newcommand{\uea}{universal enveloping algebra}
\newcommand{\lcft}{logarithmic conformal field theory}
\newcommand{\lcfts}{logarithmic conformal field theories}
\newcommand{\WZW}{Wess-Zumino-Witten}
\newcommand{\ope}{operator product expansion}
\newcommand{\opes}{operator product expansions}
\newcommand{\hws}{highest weight state}
\newcommand{\hwss}{highest weight states}
\newcommand{\hwm}{highest weight module}
\newcommand{\hwms}{highest weight modules}

\begin{document}

\title{Fusion in Fractional Level $\widehat{\mathfrak{sl}} \left( 2 \right)$-Theories with $k=-\frac{1}{2}$}

\author[D Ridout]{David Ridout}

\address[David Ridout]{
Department of Theoretical Physics \\
Research School of Physics and Engineering \\
and
Department of Mathematics \\
Mathematical Sciences Institute \\
Australian National University \\
Canberra, ACT 0200 \\
Australia
}

\email{david.ridout@anu.edu.au}

\thanks{\today}

\begin{abstract}
The fusion rules of conformal field theories admitting an $\AKMA{sl}{2}$-symmetry at level $k = -\tfrac{1}{2}$ are studied.  It is shown that the fusion closes on the set of irreducible highest weight modules and their images under spectral flow, but not when ``highest weight'' is replaced with ``relaxed highest weight''.  The fusion of the relaxed modules, necessary for a well-defined $\AKMA{u}{1}$-coset, gives two families of indecomposable modules on which the Virasoro zero-mode acts non-diagonalisably.  This confirms the logarithmic nature of the associated theories.  The structures of the indecomposable modules are completely determined as staggered modules and it is shown that there are no logarithmic couplings (beta-invariants).  The relation to the fusion ring of the $c=-2$ triplet model and the implications for the $\beta \gamma$ ghost system are briefly discussed.
\end{abstract}

\maketitle

\onehalfspacing

\section{Introduction} \label{secIntro}

This is a continuation of the study, initiated in \cite{RidSL208} and developed in \cite{RidSL210}, of the fractional level \WZW{} model based on $\AKMA{sl}{2}$ at level $k = -\tfrac{1}{2}$.  Our aim in this series of papers is to put the fractional level models on firm ground as \lcfts{} \cite{RozQua92,GurLog93}, starting with what is arguably the simplest, and perhaps most important, example.  What distinguishes this study from previous attempts, in particular that of \cite{LesSU202,LesLog04}, is the philosophy that one should use intrinsic methods wherever possible.  The resulting picture is far more complete than was previously available and we expect it to generalise in a straight-forward manner to other fractional levels.

The aim of this note is to describe, in some detail, the fusion rules of theories with $\AKMA{sl}{2}_{-1/2}$-symmetry.  In view of our stated philosophy, we will rely upon the abstract fusion algorithm developed by Nahm \cite{NahQua94} and Gaberdiel and Kausch \cite{GabInd96}.  This is described very clearly in the latter article, but see also \cite{GabInt00,GabAlg03,EbeVir06} for expositions.  This algorithm is well-suited to the exploration of theories in which one suspects representations more exotic than the irreducible highest weight ones that are familiar from rational \cft{}.  Its chief virtue is that it does not presuppose that the fusion product of two representations belongs to any given module category.  It may therefore be used to demonstrate, for example, that the category generated by the \hwms{} need not be closed under fusion (although one may have to think laterally in order to expose this).  It has so far been used to investigate module structure for the Virasoro algebra \cite{GabInd96,EbeVir06,RidPer07,RidLog07,RidPer08,RidSta09,GabFus09}, its $N=1$ and $N=2$ extensions \cite{GabFus97}, $\AKMA{sl}{2}$ at level $-\tfrac{4}{3}$ \cite{GabFus01}, and certain $\func{W}{p',p}$ algebras \cite{GabRat96,WooFus10}.

Despite its advantages, the Nahm-Gaberdiel-Kausch fusion algorithm has been criticised in the past as ``too formal'' and its application ``tedious''.  Certainly, any moderately complicated fusion process does lead to a significant amount of unpleasant algebra if done by hand, though no more so than the computation of four-point correlation functions or the \ope{} of normally-ordered products of fields.  We refer to \cite{FucNon04,PeaLog06,ReaAss07,HuaLog07,PeaInt08,KytFro08} for some alternative methods to compute fusion products.  The point is that to identify the structure of exotic representations, it is usually necessary to analyse in detail the descendant fields rather than just the primaries, and it is this that leads to the complexity.  However, the algorithm of Nahm and Gaberdiel-Kausch is straight-forward to implement within a computer algebra package, relieving a significant amount of the burden.  Our own implementation uses \textsc{Maple} and is based on a similar implementation for the Virasoro algebra.

One significant difference between the Virasoro computations and those described here for $\AKMA{sl}{2}$ is that many of the fractional level representations have an \emph{infinite} number of linearly independent states of the same conformal dimension.  However, we note that there is a definite regularity to the structure of these states.  We may therefore use symbolic calculus to encode such an infinite set of states using (rational) functions.  This is the technical realisation that we exploit in the explicit computations that follow.  We mention that the rational functions can become extremely unwieldy and that memory issues are expected to become a problem eventually.  However, the results presented here were all derived rather quickly on standard desktop workstations.

Let us briefly outline the rest of this article.  First, \secref{secOld} reviews our notations and conventions for the Kac-Moody algebra $\AKMA{sl}{2}$ and describes the results obtained in \cite{RidSL208,RidSL210} that will be needed in the sequel.  In particular, we discuss the irreducible representations that a theory with $\AKMA{sl}{2}_{-1/2}$-symmetry admits and collect explicit formulae describing the singular vectors (in the appropriate Verma-like modules) which have been set to zero in forming the irreducibles.  These ``vanishing singular vectors'' are an essential input of the Nahm-Gaberdiel-Kausch algorithm.

The algorithm itself, in its simplest form, is described in \secref{secFusion0}.  Here, we detail the explicit computations that yield the fusion rules of the irreducible \hwms{} with one another (\secref{secFusLL}) and with the irreducible \emph{relaxed} \hwms{} (\secref{secFusLE}).  Specifically, we compute what amounts to the zero-grade subspace of the fusion product and deduce the result from there.  One novel feature of this deduction is that we use the (conjectured) good behaviour of fusion under spectral flow to \emph{prove} that in each case, the fusion product involves no additional twisted modules whose presence would normally be hidden in the zero-grade analysis.  We then turn to the fusion of the relaxed \hwms{} with one another (\secref{secFusEE}), again computing just the zero-grade subspace of the result.  However, we are wary of making any deductions in this case as the proof that twisted modules do not contribute breaks down.

In fact, our wariness is justified.  In \secref{secFusion}, we revisit the fusion of the relaxed \hwms{} with one another, this time keeping track of slightly more than just the zero-grade subspace of the fusion product.  We find that the results of such fusion processes are reducible but indecomposable modules of the type referred to as \emph{staggered} modules in the Virasoro setting \cite{RohRed96,RidSta09}.  We deduce the structure of these indecomposables in terms of exact sequences (composition series) and prove that the structure uniquely specifies the module --- there are no free \emph{logarithmic couplings} in the language of \cite{RidPer07}.  This is followed by a brief account of the fusion of the relaxed \hwms{} with these new staggered indecomposables, demonstrating that the fusion ring thereby closes.  Our results are summarised in \secref{secSummary}, where we also briefly remark upon the relation between the fusion rings of $\AKMA{sl}{2}_{-1/2}$ and the $c=-2$ triplet model, and upon the implications of our results for the $\beta \gamma$ ghost system.  This summary may be read independently of the detailed fusion computations in \secDref{secFusion0}{secFusion}, although the reader will miss the explicit description of the structure of the indecomposable modules.  For this, the reader should consult \secDref{secAnalysis1}{secAnalysis2}.

Throughout, we describe the fusion algorithm and its results in significant detail in order to explain clearly how such computations are performed and to give the reader a sense of what evidence must be gathered before a conclusion is reached.  We hope that this exposition will be of use to others interested in Kac-Moody fusion beyond the integrable category.

\section{Background} \label{secOld}

We will first review the $\AKMA{sl}{2}_{-1/2}$ fractional level theories as discussed in \cite{RidSL208,RidSL210}.  We fix once and for all our preferred basis $\set{e,h,f}$ of $\SLA{sl}{2}$ to be that for which the non-trivial commutation relations are
\begin{equation}
\comm{h}{e} = 2 e, \qquad \comm{e}{f} = -h \qquad \text{and} \qquad \comm{h}{f} = -2 f.
\end{equation}
This basis is tailored to the $\SLA{sl}{2 ; \RR}$ adjoint, $e^{\dag} = f$ and $h^{\dag} = h$, and we prefer it because it is this adjoint which leads to the $\beta \gamma$ ghost system as an extended algebra of $\AKMA{sl}{2}_{-1/2}$.  The Killing form is given in this basis by
\begin{equation} \label{eqnKilling}
\killing{h}{h} = 2 \qquad \text{and} \qquad \killing{e}{f} = -1,
\end{equation}
with all other combinations giving zero.

These conventions for $\SLA{sl}{2}$ carry over to $\AKMA{sl}{2}$ in the usual way.  Replacing the central mode by $k = -\tfrac{1}{2}$ for convenience, the non-trivial commutation relations of the affine algebra are
\begin{equation}
\comm{h_m}{e_n} = 2 e_{m+n}, \quad \comm{h_m}{h_n} = -m \delta_{m+n,0}, \quad \comm{e_m}{f_n} = -h_{m+n} + \frac{1}{2} m \delta_{m+n,0} \quad \text{and} \quad \comm{h_m}{f_n} = -2 f_{m+n}.
\end{equation}
\eqnref{eqnKilling} now determines the energy-momentum tensor of the theory as
\begin{equation} \label{eqnDefT}
\func{T}{z} = \frac{1}{3} \brac{\frac{1}{2} \normord{\func{h}{z} \func{h}{z}} - \normord{\func{e}{z} \func{f}{z}} - \normord{\func{f}{z} \func{e}{z}}}.
\end{equation}
This yields the central charge $c = -1$ and a conformal dimension of $1$ for each of the primary fields $\func{e}{z}$, $\func{h}{z}$ and $\func{f}{z}$.

It is important to note that the automorphisms of $\AKMA{sl}{2}$ which preserve our choice of Cartan subalgebra are generated by the conjugation automorphism $\mathsf{w}$ and the spectral flow automorphism $\gamma$.  These act on our basis elements (with $k=-\tfrac{1}{2}$) via
\begin{subequations}
\begin{align} \label{eqnAuts}
\func{\mathsf{w}}{e_n} &= f_n, & \func{\mathsf{w}}{h_n} &= -h_n, & \func{\mathsf{w}}{f_n} &= e_n, & \func{\mathsf{w}}{L_0} &= L_0, \\
\func{\gamma}{e_n} &= e_{n-1}, & \func{\gamma}{h_n} &= h_n + \frac{1}{2} \delta_{n,0}, & \func{\gamma}{f_n} &= f_{n+1}, & \func{\gamma}{L_0} &= L_0 - \frac{1}{2} h_0 - \frac{1}{8}.
\end{align}
\end{subequations}
Moreover, they induce maps $\mathsf{w}^*$ and $\gamma^*$ from any $\AKMA{sl}{2}$-module $\mathcal{M}$ to new modules $\func{\mathsf{w}^*}{\mathcal{M}}$ and $\func{\gamma^*}{\mathcal{M}}$ (respectively).  The underlying vector spaces remain the same, but the new algebra action is given by
\begin{equation}
J \cdot \tfunc{\mathsf{w}^*}{\ket{v}} = \func{\mathsf{w}^*}{\tfunc{\mathsf{w}^{-1}}{J} \ket{v}} \qquad \text{and} \qquad J \cdot \tfunc{\gamma^*}{\ket{v}} = \func{\gamma^*}{\tfunc{\gamma^{-1}}{J} \ket{v}} \qquad \text{($J \in \AKMA{sl}{2}$).}
\end{equation}
We will not usually bother with the superscripts which distinguish the algebra automorphisms from the maps between modules.  Which is meant should be clear from the context.

There are (at least) two candidate \cfts{} with $\AKMA{sl}{2}_{-1/2}$ symmetry, distinguished by their chiral spectra.\footnote{In fact, there are most likely infinitely many, characterised as orbifolds of the $\beta \gamma$ ghost system (with maximal spectrum).}  The first is built from two infinite sequences of irreducible $\AKMA{sl}{2}$-modules $\tfunc{\gamma^{\ell}}{\AffIrrMod{0}}$ and $\tfunc{\gamma^{\ell}}{\AffIrrMod{1}}$, where $\ell \in \ZZ$.  Here, $\AffIrrMod{0}$ and $\AffIrrMod{1}$ denote the irreducible \hwms{} which are generated by \hwss{} of $\SLA{sl}{2}$-weight and conformal dimension $\brac{\lambda , h_{\lambda}} = \brac{0,0}$ and $\brac{1,\tfrac{1}{2}}$, respectively.  The former state is the vacuum of the theory.  We illustrate these families of irreducible modules schematically in \figref{figSpecFlow}.  Note that for $\abs{\ell} \geqslant 2$, the conformal dimensions of the states of the modules are no longer bounded below.

{
\psfrag{L0}[][]{$\AffIrrMod{0}$}
\psfrag{L1}[][]{$\AffIrrMod{1}$}
\psfrag{E0}[][]{$\AffOthMod{0}$}
\psfrag{E1}[][]{$\AffOthMod{1}$}
\psfrag{La}[][]{$\AffIrrMod{-1/2}$}
\psfrag{Lb}[][]{$\AffIrrMod{-3/2}$}
\psfrag{La*}[][]{$\AffIrrMod{-1/2}^*$}
\psfrag{Lb*}[][]{$\AffIrrMod{-3/2}^*$}
\psfrag{g}[][]{$\gamma$}
\psfrag{00}[][]{$\scriptstyle \brac{0,0}$}
\psfrag{aa}[][]{$\scriptstyle \brac{\tfrac{1}{2},-\tfrac{1}{8}}$}
\psfrag{bb}[][]{$\scriptstyle \brac{\tfrac{1}{2},-\tfrac{1}{8}}$}
\psfrag{cc}[][]{$\scriptstyle \brac{-1,-\tfrac{1}{2}}$}
\psfrag{dd}[][]{$\scriptstyle \brac{1,-\tfrac{1}{2}}$}
\psfrag{ee}[][]{$\scriptstyle \brac{1,\tfrac{1}{2}}$}
\psfrag{ff}[][]{$\scriptstyle \brac{-1,\tfrac{1}{2}}$}
\psfrag{gg}[][]{$\scriptstyle \brac{-\tfrac{3}{2},-\tfrac{1}{8}}$}
\psfrag{hh}[][]{$\scriptstyle \brac{\tfrac{1}{2},\tfrac{7}{8}}$}
\psfrag{ii}[][]{$\scriptstyle \brac{\tfrac{3}{2},-\tfrac{1}{8}}$}
\psfrag{jj}[][]{$\scriptstyle \brac{-\tfrac{1}{2},\tfrac{7}{8}}$}
\psfrag{kk}[][]{$\scriptstyle \brac{-2,-1}$}
\psfrag{ll}[][]{$\scriptstyle \brac{0,1}$}
\psfrag{mm}[][]{$\scriptstyle \brac{2,-1}$}
\psfrag{0e}[][]{$\scriptstyle \brac{0,-\tfrac{1}{8}}$}
\psfrag{eE}[][]{$\scriptstyle \brac{1,-\tfrac{1}{8}}$}
\psfrag{fE}[][]{$\scriptstyle \brac{-1,-\tfrac{1}{8}}$}
\psfrag{aq}[][]{$\scriptstyle \brac{-\tfrac{1}{2},-\tfrac{1}{4}}$}
\psfrag{cq}[][]{$\scriptstyle \brac{-1,-\tfrac{5}{8}}$}
\psfrag{bq}[][]{$\scriptstyle \brac{\tfrac{1}{2},-\tfrac{1}{4}}$}
\psfrag{dq}[][]{$\scriptstyle \brac{1,-\tfrac{5}{8}}$}
\psfrag{gq}[][]{$\scriptstyle \brac{-\tfrac{3}{2},-\tfrac{3}{4}}$}
\psfrag{hq}[][]{$\scriptstyle \brac{\tfrac{1}{2},\tfrac{1}{4}}$}
\psfrag{iq}[][]{$\scriptstyle \brac{\tfrac{3}{2},-\tfrac{3}{4}}$}
\psfrag{jq}[][]{$\scriptstyle \brac{-\tfrac{1}{2},\tfrac{1}{4}}$}
\psfrag{kq}[][]{$\scriptstyle \brac{-2,-\tfrac{13}{8}}$}
\psfrag{lq}[][]{$\scriptstyle \brac{0,\tfrac{3}{8}}$}
\psfrag{mq}[][]{$\scriptstyle \brac{2,-\tfrac{13}{8}}$}
\begin{figure}
\begin{center}
\includegraphics[width=\textwidth]{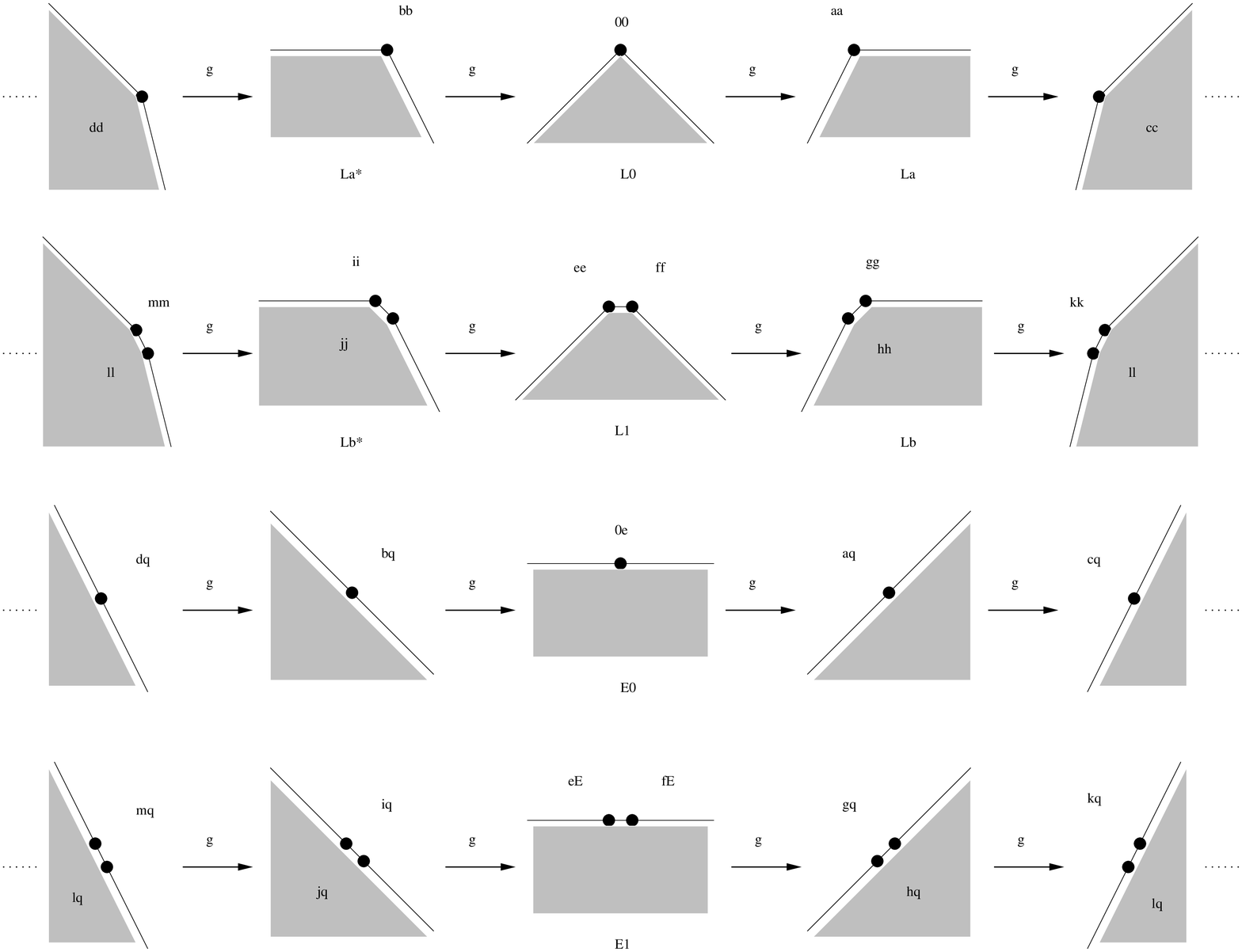}
\caption{Depictions of the modules constituting the spectra of our $\AKMA{sl}{2}_{-1/2}$-theories, emphasising the induced action of the spectral flow automorphism $\gamma$.  Each labelled state declares its $\func{\alg{sl}}{2}$-weight and conformal dimension (in that order).  Conformal dimensions increase from top to bottom and $\SLA{sl}{2}$-weights increase from right to left.} \label{figSpecFlow}
\end{center}
\end{figure}
}

We mention that $\tfunc{\gamma}{\AffIrrMod{0}}$ and $\tfunc{\gamma}{\AffIrrMod{1}}$ are also irreducible \hwms{} with respective highest weights $-\tfrac{1}{2}$ and $-\tfrac{3}{2}$.  It is therefore appropriate to write $\AffIrrMod{-1/2} = \tfunc{\gamma}{\AffIrrMod{0}}$ and $\AffIrrMod{-3/2} = \tfunc{\gamma}{\AffIrrMod{1}}$.  This brings the number of \hwms{} in the theory to four.  There are no others; in fact, these four constitute the admissible modules (for $k=-\tfrac{1}{2}$) of Kac and Wakimoto \cite{KacMod88b}.  Whereas $\AffIrrMod{0}$ and $\AffIrrMod{1}$ are both self-conjugate modules, the conjugates of $\AffIrrMod{-1/2}$ and $\AffIrrMod{-3/2}$ are the non-\hwms{} $\tfunc{\gamma^{-1}}{\AffIrrMod{0}}$ and $\tfunc{\gamma^{-1}}{\AffIrrMod{1}}$, respectively.  In general, the module conjugate to $\tfunc{\gamma^{\ell}}{\AffIrrMod{\lambda}}$, for $\lambda = 0,1$, is $\tfunc{\gamma^{-\ell}}{\AffIrrMod{\lambda}}$.

The second candidate theory extends that described above in that it is constructed from four infinite families of irreducible modules which are generated by spectral flow from the irreducibles $\AffIrrMod{0}$, $\AffIrrMod{1}$, $\AffOthMod{0}$ and $\AffOthMod{1}$.  The new modules $\AffOthMod{0}$ and $\AffOthMod{1}$ are examples of so-called \emph{relaxed} \hwms{} and are generated by \emph{relaxed} \hwss{} \cite{FeiEqu98,SemEmb97}.  These are states that would be genuine \hwss{} except for the fact that they need not be annihilated by $e_0$.  It is not hard to see that every zero-grade state $\ket{v_m}$ of $\AffOthMod{0}$ and $\AffOthMod{1}$ is a relaxed \hws{}.  The common conformal dimension of these zero-grade states is $-\tfrac{1}{8}$ and their $\SLA{sl}{2}$-weights $m$ are either even ($\AffOthMod{0}$) or odd ($\AffOthMod{1}$).  Moreover, there is a single $\ket{v_m}$ (up to scalar multiples) for each weight $m$ and they are related by the $\AKMA{sl}{2}$-action as follows:
\begin{equation} \label{eqnEModConv}
e_0 \ket{v_m} = \ket{v_{m + 2}} \qquad \text{and} \qquad f_0 \ket{v_m} = \frac{\brac{2 m - 1} \brac{2 m - 3}}{16} \ket{v_{m - 2}}.
\end{equation}
The $\AffOthMod{\lambda}$ and their images under spectral flow are also illustrated schematically in \figref{figSpecFlow}.  We note that $\tfunc{\gamma^{\ell}}{\AffOthMod{\lambda}}$ and $\tfunc{\gamma^{-\ell}}{\AffOthMod{\lambda}}$ are conjugate modules and that the conformal dimensions of the states of $\tfunc{\gamma^{\ell}}{\AffOthMod{\lambda}}$ are not bounded below when $\abs{\ell} \geqslant 1$.

The irreducible module $\AffIrrMod{1}$ plays a special role in these theories because it fuses with itself to give the vacuum module $\AffIrrMod{0}$:
\begin{equation} \label{eqnFusL1xL1}
\AffIrrMod{1} \fuse \AffIrrMod{1} = \AffIrrMod{0}.
\end{equation}
This was argued to be true in \cite{RidSL208} and we shall give a full proof in \secref{secFusLL}.  This property makes $\AffIrrMod{1}$ an order $2$ simple current by which we may extend the chiral algebra $\AKMA{sl}{2}_{-1/2}$ of our theories.  The resulting extended chiral algebra is the well known $\beta \gamma$ ghost algebra.  This is a free field algebra that is generated by two bosonic fields, $\beta$ and $\gamma$, of dimension $\tfrac{1}{2}$ whose modes satisfy $\beta_n^{\dag} = \gamma_{-n}$,
\begin{equation} \label{eqnGhostComm}
\comm{\beta_m}{\beta_n} = \comm{\gamma_m}{\gamma_n} = 0 \qquad \text{and} \qquad \comm{\gamma_m}{\beta_n} = \delta_{m+n,0}.
\end{equation}
Chirally, the modules $\AffIrrMod{0}$ and $\AffIrrMod{1}$ combine to form a single irreducible module for the $\beta \gamma$ ghost algebra.  As we shall see in \secref{secFusLE}, so too do $\AffOthMod{0}$ and $\AffOthMod{1}$.  Moreover, both conjugation and spectral flow lift to automorphisms of the extended algebra, so we end up with two infinite families of irreducible $\beta \gamma$-modules, related by spectral flow (an $\AffIrrMod{}$-type and an $\AffOthMod{}$-type family).

The fusion rules of the $\AffIrrMod{\lambda}$ are then derived from \eqnref{eqnFusL1xL1} by using the following formulae, assumed to be valid for all modules $\mathcal{M}$ and $\mathcal{N}$:
\begin{equation} \label{eqnFusionAssumption}
\func{\mathsf{w}}{\mathcal{M}} \fuse \func{\mathsf{w}}{\mathcal{N}} = \func{\mathsf{w}}{\mathcal{M} \fuse \mathcal{N}} \qquad \text{and} \qquad \func{\gamma^{\ell_1}}{\mathcal{M}} \fuse \func{\gamma^{\ell_2}}{\mathcal{N}} = \func{\gamma^{\ell_1 + \ell_2}}{\mathcal{M} \fuse \mathcal{N}}.
\end{equation}
The first is in fact not difficult to prove, but we know of no proof for the second despite much evidence in its favour.  We mention however that the second formula does hold for the integrable modules of the rational \WZW{} models, though the standard proof is far from elementary (it relies upon the Verlinde formula --- see for example \cite[Sec.~16.1]{DiFCon97}).

It has long been known that the characters of the four admissible \hwms{} $\AffIrrMod{0}$, $\AffIrrMod{1}$, $\AffIrrMod{-1/2}$ and $\AffIrrMod{-3/2}$ close under the usual action of the modular group.  Because the latter two are spectral flow images of the former two, we learn from \eqref{eqnFusionAssumption} that the fusion rules do not close on these four modules.  In fact, the smallest set of modules containing these four which is closed under fusion and conjugation consists precisely of the two infinite families which constitute the spectrum of the first $\AKMA{sl}{2}_{-1/2}$-theory discussed above.  Moreover, the characters of this spectrum still form a four-dimensional representation of the modular group due to certain periodicities in the characters under spectral flow.  For a \emph{rational} theory, this closure under fusion and modular invariance would be taken as strong evidence that one can construct a consistent \cft{} from this spectrum.  However, the theory is not rational because of the infinite number of distinct modules.

The second $\AKMA{sl}{2}_{-1/2}$-theory discussed above can be motivated by the observation that the coset theory of the first by the $\AKMA{u}{1}$-subtheory generated by the field $h$ is not modular invariant.  Indeed, this coset gives only two of the four irreducible modules which can be regarded as the building blocks of that archetype of \lcft{}, the $c=-2$ triplet model \cite{GabRat96}.  In order to obtain the remaining two irreducibles, the spectrum of the first $\AKMA{sl}{2}_{-1/2}$-theory must be augmented by $\AffOthMod{0}$ and $\AffOthMod{1}$.  Invariance under spectral flow and conjugation then leads to the four families of irreducibles that generate our second $\AKMA{sl}{2}_{-1/2}$-theory.  This augmentation even preserves modular invariance, although in a somewhat weaker sense than one would like \cite{RidSL210}.

In contrast to the first theory, we do not expect that the fusion rules of our second $\AKMA{sl}{2}_{-1/2}$-theory close on the irreducibles $\AffIrrMod{0}$, $\AffIrrMod{1}$, $\AffOthMod{0}$ and $\AffOthMod{1}$ (and their images under spectral flow).  Indeed, the two additional irreducible triplet modules which necessitated the augmentation of our spectrum are known to fuse into indecomposable modules, giving the triplet model its logarithmic structure.  We therefore expect that fusing the $\AffOthMod{\lambda}$ with one another will also lead to indecomposables.  Verifying this, and analysing the resulting logarithmic structure, is in fact the main aim of what follows.

It remains to collect some explicit formulae which will be useful in achieving this aim.  To compute the fusion rules involving $\AffIrrMod{0}$, $\AffIrrMod{1}$, $\AffOthMod{0}$ and $\AffOthMod{1}$, we will make use of explicit expressions for the (relaxed) singular vectors of the (relaxed) Verma modules that have been set to zero upon forming the irreducible quotients.  Such quotients yield non-trivial relations which give rise to so-called \emph{spurious states} when computing fusion products \cite{NahQua94,GabInd96}.  Setting the non-trivial vacuum singular vector to zero in $\AffIrrMod{0}$ gives
\begin{equation} \label{eqnSVL0}
\brac{156 e_{-3} e_{-1} - 71 e_{-2}^2 + 44 e_{-2} h_{-1} e_{-1} - 52 h_{-2} e_{-1}^2 + 16 f_{-1} e_{-1}^3 - 4 h_{-1}^2 e_{-1}^2} \ket{0} = 0
\end{equation}
and repeating this for $\AffIrrMod{1}$ yields
\begin{equation} \label{eqnSVL1}
\brac{7 e_{-2} - 2 h_{-1} e_{-1}} \ket{u_1} + 4 e_{-1}^2 \ket{u_{-1}} = 0.
\end{equation}
Here, $\ket{u_1}$ and $\ket{u_{-1}} = f_0 \ket{u_1}$ denote the two zero-grade states of $\AffIrrMod{1}$.  To obtain $\AffOthMod{0}$ and $\AffOthMod{1}$, one has to quotient the corresponding relaxed Verma modules by submodules which are themselves relaxed Verma modules.  In this case we do not have a single generating singular vector, but rather two infinite families of relaxed singular vectors (relaxed \hwss{}).  Happily, these have a regular explicit form for both $\AffOthMod{0}$ and $\AffOthMod{1}$.  At grade one, the relaxed singular vectors give the relations
\begin{equation} \label{eqnSV1E}
\frac{\brac{2 m - 1} \brac{2 m + 3}}{16} e_{-1} \ket{v_{m - 2}} - \frac{2 m + 3}{4} h_{-1} \ket{v_m} + f_{-1} \ket{v_{m + 2}} = 0
\end{equation}
and at grade two we obtain the independent relations
\begin{multline} \label{eqnSV2E}
\frac{\brac{2m-7} \brac{2m-3} \brac{2m+1} \brac{2m+5}}{256} e_{-1}^2 \ket{v_{m-4}} - \frac{\brac{2m-3} \brac{2m+1} \brac{2m+5}}{32} \brac{h_{-1} e_{-1} - e_{-2}} \ket{v_{m-2}} \\
+ \frac{\brac{2m+1} \brac{2m+5}}{16} \brac{h_{-1}^2 + 2 f_{-1} e_{-1} - h_{-2}} \ket{v_m} - \frac{2m+5}{2} \brac{f_{-1} h_{-1} - f_{-2}} \ket{v_{m + 2}} + f_{-1}^2 \ket{v_{m + 4}} = 0.
\end{multline}
These relations appear somewhat asymmetric, but this is because we have chosen to relate the zero-grade states $\ket{v_m}$ as in \eqnref{eqnEModConv}.  Substituting this back into the above relations leads to more symmetric forms.  We find the latter forms useful when applying spectral flow to the above vanishing vectors.

Let us mention that the non-trivial vacuum relation \eqref{eqnSVL0} leads, in the usual way \cite{FeiAnn92}, to non-trivial constraints on the spectra of our $\AKMA{sl}{2}_{-1/2}$-theories.  In particular, the $\SLA{sl}{2}$-weight of any \hws{} is restricted to being $0$, $1$, $-\tfrac{1}{2}$ or $-\tfrac{3}{2}$.  It follows that the only \hwms{} that can be admitted in the theory are the four irreducibles $\AffIrrMod{0}$, $\AffIrrMod{1}$, $\AffIrrMod{-1/2}$ and $\AffIrrMod{-3/2}$.  Similarly, relaxed \hwss{} are restricted to either being a zero-grade state of $\AffIrrMod{0}$ or $\AffIrrMod{1}$, or having conformal dimension $-\tfrac{1}{8}$.  This covers all the zero-grade states of $\AffIrrMod{-1/2}$ and $\AffIrrMod{-3/2}$, their conjugates $\AffIrrMod{-1/2}^* = \tfunc{\gamma^{-1}}{\AffIrrMod{0}}$ and $\AffIrrMod{-3/2}^* = \tfunc{\gamma^{-1}}{\AffIrrMod{0}}$, as well as the zero-grade states of $\AffOthMod{0}$ and $\AffOthMod{1}$.

However, it also allows more general modules $\AffOthMod{\lambda}$ with $\lambda \notin \ZZ$, provided that their zero-grade states $\ket{v_m}$ have $\SLA{sl}{2}$-weight $m$ and satisfy \eqnref{eqnEModConv}.  Of course, we have the identification $\AffOthMod{\lambda} = \AffOthMod{\lambda+2}$.  If $\lambda \notin \ZZ + \tfrac{1}{2}$, then $\AffOthMod{\lambda}$ is irreducible and the relations \eqref{eqnSV1E} and \eqref{eqnSV2E} still hold.  The case $\lambda \in \ZZ + \tfrac{1}{2}$ is interesting as the relation \eqref{eqnSVL0} admits four distinct indecomposable relaxed \hwms{}, two with lowest weight states of $\SLA{sl}{2}$-weights $\tfrac{1}{2}$ and $\tfrac{3}{2}$, and two with \hwss{} of $\SLA{sl}{2}$-weights $-\tfrac{1}{2}$ and $-\tfrac{3}{2}$.  We will denote these indecomposables by $\AffOthMod{1/2}^-$, $\AffOthMod{3/2}^-$, $\AffOthMod{-1/2}^+$ and $\AffOthMod{-3/2}^+$, respectively, noting that conjugation gives
\begin{equation}
\func{\mathsf{w}}{\AffOthMod{1/2}^-} = \AffOthMod{-1/2}^+ \qquad \text{and} \qquad \func{\mathsf{w}}{\AffOthMod{3/2}^-} = \AffOthMod{-3/2}^+.
\end{equation}
The zero-grade states of $\AffOthMod{1/2}^-$ and $\AffOthMod{3/2}^-$ still satisfy \eqnref{eqnEModConv}, though those of $\AffOthMod{-1/2}^+$ and $\AffOthMod{-3/2}^+$ will not --- \eqref{eqnEModConv} manifestly assumes no \hwss{}.  Rather, the states of $\AffOthMod{-1/2}^+$ and $\AffOthMod{-3/2}^+$ may be taken to satisfy the equations obtained by applying $\mathsf{w}$ to \eqref{eqnEModConv}.

\section{Fusion to Grade $0$} \label{secFusion0}

\subsection{Preliminaries} \label{secFus0Pre}

We now turn to the fusion rules of the irreducible modules $\AffIrrMod{0}$, $\AffIrrMod{1}$, $\AffOthMod{0}$ and $\AffOthMod{1}$.  These will be calculated with the help of the algorithm of Nahm and Gaberdiel-Kausch \cite{NahQua94,GabInd96} which abstracts, in terms of coproduct formulae, the natural action(s) of the chiral symmetry algebra on the chiral \opes{} of the theory.  The key assumption underlying this algorithm is that the vector space of the fusion product of two modules may be realised as a quotient (subspace) of that of the (vector space) tensor product of these modules.  Given this, the master formulae defining the fusion coproduct for affine Kac-Moody algebras are most usefully given in the forms \cite{GabFus94}
\begin{subequations} \label{eqnMaster}
\begin{align}
\func{\Delta}{J_n} &= \sum_{m=0}^n \binom{n}{m} J_m \otimes \wun + \wun \otimes J_n & \text{(} n &\geqslant 0 \text{)}  \label{eqnMaster1} \\
\func{\Delta}{J_{-n}} &= \sum_{m=0}^{\infty} \binom{n+m-1}{m} \brac{-1}^m J_m \otimes \wun + \wun \otimes J_{-n} & \text{(} n &\geqslant 1 \text{)}  \label{eqnMaster2} \\
J_{-n} \otimes \wun &= \sum_{m=n}^{\infty} \binom{m-1}{n-1} \func{\Delta}{J_{-m}} - \brac{-1}^n \sum_{m=0}^{\infty} \binom{n+m-1}{m} \wun \otimes J_m & \text{(} n &\geqslant 1 \text{).} \label{eqnMaster3}
\end{align}
\end{subequations}
Here, $\otimes$ denotes the usual vector space tensor product (over $\CC$) and $J$ stands for either $e$, $h$ or $f$.  The first two formulae define the action of the affine modes on the fusion module, whereas the last may be viewed as a necessary auxiliary formula for explicit computation (it actually amounts to imposing the equivalence of two distinct fusion coproducts).  We will also need the coproduct formula for the Virasoro zero-mode:
\begin{equation} \label{eqnMaster4}
\func{\Delta}{L_0} = L_{-1} \otimes \wun + L_0 \otimes \wun + \wun \otimes L_0.
\end{equation}

Note that for these sums appearing in \eqref{eqnMaster} to be finite, the modules to be fused should have their subspaces of constant $\SLA{sl}{2}$-weight consist of states whose conformal dimensions are bounded below.  This is the case for all the modules that we shall consider as this property is preserved by the induced action of the spectral flow automorphism $\gamma$.  Even so, the first sum in \eqnref{eqnMaster3} will still be infinite.  However, we will only be interested in computing in certain quotients of the modules, and this will truncate the remaining infinite sum.  In this section, we will restrict ourselves to explicitly computing only the most readily available information about the fusion rules.  We refer to this as fusing to grade $0$ because from this we will only obtain information about the zero-grade states of the fusion module.

To compute this grade $0$ fusion of two $\AKMA{sl}{2}$-modules, one applies these formulae in the (vector space) tensor product of the zero-grade subspaces of both modules.  In general, vectors which vanish in either module, but not in their Verma or Verma-like parents, will induce linear relations in this tensor product space which must be imposed to get the correct fusion space.  Such linear relations are referred to as spurious states \cite{NahQua94}.  We mention that the vanishing vectors which give rise to the spurious states do not have to belong to the zero-grade subspaces.

Before beginning the calculations, it will be useful to examine the basic premise of the fusion algorithm in slightly more detail.  This somewhat formal discussion makes the above description precise and makes contact with the generalisations necessary for fusing beyond grade $0$ (\secref{secFusion}).  Let us define $\alg{A}^-$ to be the subalgebra of the \uea{} of $\AKMA{sl}{2}$ which is generated by the $e_{-n}$, $h_{-n}$ and $f_{-n}$ with $n \geqslant 1$.  This obviously acts on $\AKMA{sl}{2}$-modules.  A precise version of the above claim regarding the fusion of the $\AKMA{sl}{2}$-modules $\mathcal{M}$ and $\mathcal{N}$ to grade $0$ is then that
\begin{equation} \label{eqnFusionGrade0}
\frac{\mathcal{M} \fuse \mathcal{N}}{\func{\func{\Delta}{\alg{A}^-}}{\mathcal{M} \fuse \mathcal{N}}} \subseteq \frac{\mathcal{M}}{\alg{A}^- \mathcal{M}} \otimes \frac{\mathcal{N}}{\alg{A}^- \mathcal{N}}
\end{equation}
as (complex) vector spaces.  The point here is that $\mathcal{M} / \alg{A}^- \mathcal{M}$ reduces to the usual notion of zero-grade subspace when $\mathcal{M}$ is a (relaxed) \hwm{}.  Then, we can interpret this relation as saying that the zero-grade subspace of the fusion product may be found within the tensor product of the zero-grade subspaces of the original modules.  However, \eqref{eqnFusionGrade0} is a generalisation of this which makes sense for all modules $\mathcal{M}$, in particular for the images of (relaxed) \hwms{} under spectral flow.

Proving \eqref{eqnFusionGrade0} amounts to demonstrating that the following procedure terminates.  Consider a representative state $\ket{v} \otimes \ket{w} \in \mathcal{M} \otimes \mathcal{N}$ for an element of the left hand side of \eqref{eqnFusionGrade0}.
\begin{enumerate}
\item \label{Step1} If $\ket{v} = J_{-n} \ket{v'}$ for some $J_{-n} \in \alg{A}^-$, then we apply \eqnref{eqnMaster3} to obtain
\begin{equation} \label{eqnStep1}
\ket{v} \otimes \ket{w} = -\brac{-1}^n \sum_{m=0}^{\infty} \binom{n+m-1}{m} \ket{v'} \otimes J_m \ket{w},
\end{equation}
as for $m \geqslant 1$, $\func{\Delta}{J_{-m}} = 0$ when acting upon the left hand side of \eqref{eqnFusionGrade0}.
\item \label{Step2} If $\ket{w} = J_{-n} \ket{w'}$ for some $J_{-n} \in \alg{A}^-$, then we apply \eqnref{eqnMaster2} to $\func{\func{\Delta}{J_{-n}}}{\ket{v} \otimes \ket{w'}} = 0$, obtaining
\begin{equation} \label{eqnStep2}
\ket{v} \otimes \ket{w} = -\sum_{m=0}^{\infty} \binom{n+m-1}{m} \brac{-1}^m J_m \ket{v} \otimes \ket{w'}.
\end{equation}
\end{enumerate}
Repeating these steps as needed, any representative state for the left hand side of \eqref{eqnFusionGrade0} should be reduced to a linear combination of representative states for the right hand side.

However, the actual termination of this procedure is not \emph{a priori} guaranteed.  In step \ref{Step1}, $\ket{v'}$ might have the form $J_{-n'} \ket{v''}$ with $J_{-n'} \in \alg{A}^-$, so we would have to apply step \ref{Step1} again.  We thereby see that this part of the procedure will terminate if every $\ket{v}$ (from an appropriate basis) has the form $J_{-n_1} J_{-n_2} \cdots J_{-n_t} \ket{u}$ for some $\ket{u} \in \mathcal{M} / \alg{A}^- \mathcal{M}$.  One can check that for $\mathcal{M}$ of the form $\AffIrrMod{\lambda}$, $\AffOthMod{\lambda}$ or $\tfunc{\gamma^{\pm 1}}{\AffIrrMod{\lambda}}$, this is guaranteed, hence termination is inevitable.  The analysis is identical for step \ref{Step2}, so we conclude that when both $\mathcal{M}$ and $\mathcal{N}$ are of the form $\AffIrrMod{\lambda}$, $\AffOthMod{\lambda}$ or $\tfunc{\gamma^{\pm 1}}{\AffIrrMod{\lambda}}$, then the (grade $0$) fusion algorithm terminates.

In the remaining cases, when either $\mathcal{M}$ or $\mathcal{N}$ is one of the twisted modules $\tfunc{\gamma^{\ell}}{\AffIrrMod{\lambda}}$ with $\abs{\ell} > 1$ or $\tfunc{\gamma^{\ell}}{\AffOthMod{\lambda}}$ with $\ell \neq 0$, one can check that the respective quotient $\mathcal{M} / \alg{A}^- \mathcal{M}$ or $\mathcal{N} / \alg{A}^- \mathcal{N}$ is in fact trivial.  Termination is therefore not clear, and in fact seems rather unlikely.  Worse yet, applying step \ref{Step2} might lead to new states to which step \ref{Step1} should be applied and vice-versa.  We conclude that the termination of the fusion algorithm is a subtle business in general, even when computing to grade $0$.

Finally, we mention that a lack of termination does not necessarily mean that one cannot use the fusion algorithm at all.  Rather, it means that \eqnref{eqnFusionGrade0} is not appropriate for the modules which one is trying to fuse, and an alternative space must be sought for the right hand side.  We shall see an example of this in \secref{secFusES}.  In what follows, we shall take some care to consider the termination of the fusion algorithm wherever possible.

\subsection{Fusing $\AffIrrMod{1}$ and $\AffIrrMod{1}$} \label{secFusLL}

We begin by investigating the fusion of the irreducible $\AKMA{sl}{2}$-module $\AffIrrMod{1}$ with itself.  We have already given the result in \secref{secOld}, but it serves the illustrate the fusion procedure in a very simple setting, while paying close attention to the subtleties that one has to deal with in affine theories.  Letting $\ket{u_1}$ and $\ket{u_{-1}} = f_0 \ket{u_1}$ denote the zero-grade states of $\AffIrrMod{1}$, the fusion to grade $0$ will be contained within the space spanned by
\begin{equation}
\ket{u_1} \otimes \ket{u_1}, \qquad \ket{u_1} \otimes \ket{u_{-1}}, \qquad \ket{u_{-1}} \otimes \ket{u_1} \qquad \text{and} \qquad \ket{u_{-1}} \otimes \ket{u_{-1}}.
\end{equation}
This follows from the above termination discussion:  Both step \ref{Step1} and step \ref{Step2} are guaranteed to terminate, and it is easily checked that we do not need to apply step \ref{Step1} again after completing step \ref{Step2}.

Note that the $\SLA{sl}{2}$-weights of the spanning states are $2$, $0$, $0$ and $-2$, so the weight spaces have dimension $1$ or $2$.  This is well-defined because the $\SLA{sl}{2}$-weight is conserved by the fusion operation (as one expects from \opes{}).  This follows readily from taking $n=0$ in \eqnref{eqnMaster1} to get the usual tensor coproduct\footnote{It also follows from this formula that fusing two modules on which $h_0$ is diagonalisable, \hwms{} for instance, will result in a module on which $h_0$ is diagonalisable.  This means that the logarithmic \cfts{} that we are generating will have the affine zero-mode acting semisimply.  This argument does not apply to $L_0$ as \eqnref{eqnMaster4} shows.}
\begin{equation} \label{eqnMaster0}
\func{\Delta}{J_0} = J_0 \otimes \wun + \wun \otimes J_0,
\end{equation}
with $J = e$, $h$ or $f$.

Let us first remark that substituting the Sugawara form of $L_{-1}$ into \eqnref{eqnMaster4} and applying \eqnref{eqnMaster3} gives
\begin{equation} \label{eqnDeltaL0Grade0}
\func{\Delta}{L_0} = L_0 \otimes \wun + \wun \otimes L_0 + \frac{1}{3} h_0 \otimes h_0 - \frac{2}{3} e_0 \otimes f_0 - \frac{2}{3} f_0 \otimes e_0
\end{equation}
on the zero-grade subspace of $\AffIrrMod{1} \fuse \AffIrrMod{1}$.  We therefore find that $L_0$ is represented on the $\SLA{sl}{2}$-weight spaces of weights $2$, $0$ and $-2$ by
\begin{equation}
\func{\Delta}{L_0} = \frac{4}{3}, \qquad \func{\Delta}{L_0} = 
\begin{pmatrix}
\frac{2}{3} & \frac{2}{3} \\
\frac{2}{3} & \frac{2}{3}
\end{pmatrix}
\qquad \text{and} \qquad \func{\Delta}{L_0} = \frac{4}{3},
\end{equation}
respectively.  The matrix in the middle is diagonalisable with eigenvalues $0$ and $\tfrac{4}{3}$, so we have an $\SLA{sl}{2}$ singlet of dimension $0$ and a dimension $\tfrac{4}{3}$ triplet.  However, zero-grade states are forbidden from having the latter conformal dimension (\secref{secOld}), so we conclude that only the eigenstate of dimension $0$ is actually present in the fusion.  The rest must be set to zero (they must provide examples of spurious states).

This is very encouraging, but we will take some time to re-analyse the situation using the more rigorous algorithmic approach.  In part, this serves to illustrate the general procedure, which can become quite involved, but it also serves to allay doubts that the above argument might have loopholes.  In particular, one might imagine that the ``forbidden eigenstates'' of dimension $\tfrac{4}{3}$ might belong to some peculiar indecomposable module for which the dimension argument of \secref{secOld} does not apply.

We therefore turn to the vanishing vectors of the second copy of $\AffIrrMod{1}$ in order to deduce relations between the states of the weight spaces.  Such vectors are descended from the (vanishing) singular vector
\begin{equation} \label{eqnL1SV}
\brac{7 e_{-2} - 2 h_{-1} e_{-1}} \ket{u_1} + 4 e_{-1}^2 \ket{u_{-1}} = 0.
\end{equation}
As we are computing to grade $0$, $\func{\Delta}{J_{-n}}$ must be identically zero for each $J = e,h,f$ and all $n \geqslant 1$.  Thus, \eqnref{eqnMaster2} gives
\begin{subequations} \label{eqnLLSpurious}
\begin{align}
0 &= \func{\Delta}{7 e_{-2}} \ket{u} \otimes \ket{u_1} = 7 e_0 \ket{u} \otimes \ket{u_1} + 7 \ket{u} \otimes e_{-2} \ket{u_1}, \\
0 &= \func{\Delta}{-2 h_{-1} e_{-1}} \ket{u} \otimes \ket{u_1} \notag \\
& \mspace{0mu} = -2 h_0 e_0 \ket{u} \otimes \ket{u_1} - 2 e_0 \ket{u} \otimes h_{-1} \ket{u_1} - 2 h_0 \ket{u} \otimes e_{-1} \ket{u_1} - 2 \ket{u} \otimes h_{-1} e_{-1} \ket{u_1}, \\
0 &= \func{\Delta}{4 e_{-1}^2} \ket{u} \otimes \ket{u_{-1}} = 4 e_0^2 \ket{u} \otimes \ket{u_{-1}} + 8 e_0 \ket{u} \otimes e_{-1} \ket{u_{-1}} + 4 \ket{u} \otimes e_{-1}^2 \ket{u_{-1}},
\end{align}
\end{subequations}
where $\ket{u}$ might be $\ket{u_1}$ or $\ket{u_{-1}}$.  For $\ket{u} = \ket{u_1}$, we add these results and take into account the vanishing singular vector \eqref{eqnL1SV} to get
\begin{equation}
-2 \ket{u_1} \otimes e_{-1} \ket{u_1} = 0.
\end{equation}
We therefore apply step \ref{Step2} of the fusion algorithm to rewrite the left hand side as
\begin{equation}
2 e_0 \ket{u_1} \otimes \ket{u_1},
\end{equation}
which vanishes identically.  This means that no spurious states are obtained.  However, when $\ket{u} = \ket{u_{-1}}$, repeating this computation gives
\begin{equation}
-5 \ket{u_1} \otimes \ket{u_1} + 2 \ket{u_1} \otimes h_{-1} \ket{u_1} + 2 \ket{u_{-1}} \otimes e_{-1} \ket{u_1} -8 \ket{u_1} \otimes e_{-1} \ket{u_{-1}} = 0,
\end{equation}
and applying step \ref{Step2} to the left hand side now gives
\begin{equation}
-5 \ket{u_1} \otimes \ket{u_1} = 0.
\end{equation}
It follows that $\ket{u_1} \otimes \ket{u_1}$ is a spurious state, so the weight $2$ space is in fact trivial.

We can deduce further spurious states from this one by applying $\func{\Delta}{e_0}$ and $\func{\Delta}{f_0}$.  In this way, we obtain
\begin{equation}
\ket{u_1} \otimes \ket{u_{-1}} + \ket{u_{-1}} \otimes \ket{u_1} = 0 \qquad \text{and} \qquad \ket{u_{-1}} \otimes \ket{u_{-1}} = 0.
\end{equation}
We get no further spurious states by using descendants of the singular vector \eqref{eqnL1SV}, nor by using the vanishing singular vector of the first copy of $\AffIrrMod{1}$ (using \eqnref{eqnMaster3} and step \ref{Step1}), so we conclude\footnote{In fact, it is difficult to ever be sure that the relations derived are exhaustive.  However, in practice the module structure one deduces from an incomplete set of relations is almost always found to be inconsistent (especially when one computes beyond grade $0$).} that the fusion to grade $0$ is one-dimensional.  The surviving weight space has weight $0$ and one can check from \eqnref{eqnDeltaL0Grade0} that the corresponding conformal dimension is indeed $0$.

The obvious conclusion to draw from this is that
\begin{equation} \label{eqnL1xL1=L0}
\AffIrrMod{1} \fuse \AffIrrMod{1} = \AffIrrMod{0}
\end{equation}
as $\AffIrrMod{0}$ is the only admissible module with this zero-grade subspace.  This is what was reported in \cite{RidSL208} (and \secref{secOld}).  However, we should be careful and note that the computations we have carried out will not be sensitive to modules whose zero-grade subspace is trivial.  As we have already noted, these include those twisted modules of our theory whose conformal dimensions are not bounded below.  This means that it is possible that modules such as $\tfunc{\gamma^{\ell}}{\AffIrrMod{\lambda}}$ ($\abs{\ell} > 1$) and $\tfunc{\gamma^{\ell}}{\AffOthMod{\lambda}}$ ($\ell \neq 0$) could contribute to the decomposition of the fusion process, and the above computations will not see them.

To investigate the possible appearance of twisted modules, we should try to repeat our computations beyond the zeroth grade.  More precisely, this entails replacing the algebra $\alg{A}^-$ in the fusion algorithm of \secref{secFus0Pre} by a subalgebra which will detect twisted modules.  However, we are already assuming that fusion respects the spectral flow as in \eqnref{eqnFusionAssumption}, so it turns out that there is a second, easier, path which we can take.

Assume then that the fusion rule \eqref{eqnL1xL1=L0} is not correct, because there are states on the right hand side that are associated to twisted modules.  We may choose a candidate twisted module, $\tfunc{\gamma^{\ell}}{\mathcal{M}}$ say, and test for its presence in the fusion by considering instead the fusion
\begin{equation} \label{eqnL1xTwistL1}
\AffIrrMod{1} \fuse \tfunc{\gamma^{-\ell}}{\AffIrrMod{1}}.
\end{equation}
By \eqnref{eqnFusionAssumption}, if our chosen twisted module appears in \eqref{eqnL1xL1=L0}, then its \emph{untwisted} version $\mathcal{M}$ will appear in \eqref{eqnL1xTwistL1}.  The fusion algorithm of \secref{secFus0Pre} will now detect $\mathcal{M}$, provided of course that the algorithm terminates when applied to \eqref{eqnL1xTwistL1}.

We therefore examine the termination of the fusion algorithm applied to a state $\ket{v} \otimes \ket{w} \in \AffIrrMod{1} \otimes \tfunc{\gamma^{-\ell}}{\AffIrrMod{1}}$.  Step \ref{Step1} obviously still terminates, so we may assume that $\ket{v}$ is $\ket{u_1}$ or $\ket{u_{-1}}$.  Iterating step \ref{Step2} then allows us to assume that $\ket{w}$ is a state of minimal conformal dimension for its $\SLA{sl}{2}$-weight.  If the twist parameter $\ell$ has $\abs{\ell} = 1$, then we have already shown that the algorithm terminates (\secref{secFus0Pre}) for \eqref{eqnL1xTwistL1}.  For $\abs{\ell} > 1$, we may define an infinite sequence of states $\ket{w_i} \in \tfunc{\gamma^{-\ell}}{\AffIrrMod{1}}$ by
\begin{equation} \label{eqnTheSequence}
\ket{w} = J_{-n_1} \ket{w_1} = J_{-n_1} J_{-n_2} \ket{w_2} = J_{-n_1} J_{-n_2} J_{-n_3} \ket{w_3} = \cdots,
\end{equation}
in which each $\ket{w_i}$ also has the minimal conformal dimension for its $\SLA{sl}{2}$-weight.  Here, $J$ denotes either $f$ or $e$ according as to whether $\ell$ is positive or negative.  But we can only apply step \ref{Step2} to $\ket{v} \otimes \ket{w}$ twice before $\ket{v}$ is annihilated (by $J_0^2$).  As this application introduces no states to which step \ref{Step1} must be applied, the fusion algorithm thereby terminates.

However, for $\abs{\ell} > 1$, $\tfunc{\gamma^{-\ell}}{\AffIrrMod{1}}$ has trivial zero-grade subspace, hence the fusion product must be trivial by \eqnref{eqnFusionGrade0}.  It follows that $\mathcal{M}$ does not appear in the fusion \eqref{eqnL1xTwistL1}, hence that $\tfunc{\gamma^{\ell}}{\mathcal{M}}$ cannot appear in \eqref{eqnL1xL1=L0}.  For $\abs{\ell} = 1$, $\mathcal{M}$ must be of the form $\AffOthMod{\lambda}$ for $\tfunc{\gamma^{\ell}}{\mathcal{M}}$ to be undetectable in \eqref{eqnL1xL1=L0}.  But, the weights of the zero-grade subspace of $\tfunc{\gamma^{-\ell}}{\AffIrrMod{1}}$ are bounded either above or below, so those of the (vector space) tensor product of the zero-grade subspaces of $\AffIrrMod{1}$ and $\tfunc{\gamma^{-\ell}}{\AffIrrMod{1}}$ are similarly bounded.  It is now clear that $\AffOthMod{\lambda}$ cannot appear in the fusion \eqref{eqnL1xTwistL1} because the weights of its zero-grade subspace are neither bounded above nor below --- such an appearance would contradict \eqnref{eqnFusionGrade0}.  We therefore conclude that \eqnref{eqnL1xTwistL1} is indeed correct after all.  There are no contributions to the right hand side associated with unseen (to grade $0$) twisted modules.

\subsection{Fusing $\AffIrrMod{1}$ and $\AffOthMod{\lambda}$} \label{secFusLE}

We can now turn to the elucidation of new fusion rules, in particular to the fusion of the irreducible $\AKMA{sl}{2}$-modules $\AffIrrMod{1}$ and $\AffOthMod{\lambda}$.  Letting $\ket{u_1}$ and $\ket{u_{-1}} = f_0 \ket{u_1}$ denote the zero-grade states of $\AffIrrMod{1}$ and $\ket{v_m}$, $m \in 2 \ZZ + \lambda$, denote those of $\AffOthMod{\lambda}$ (normalised as in \eqnref{eqnEModConv}), the result of this grade $0$ fusion will be contained within the space spanned by
\begin{equation}
\ket{u_1} \otimes \ket{v_m} \qquad \text{and} \qquad \ket{u_{-1}} \otimes \ket{v_{m+2}} \qquad \text{($m \in 2 \ZZ + \lambda$).}
\end{equation}
In contrast to \secref{secFusLL}, this is an infinite-dimensional space.  However, the weight spaces are only two-dimensional, so we can still invoke linear algebra on these spaces separately.

Note first that applying \eqnref{eqnDeltaL0Grade0} to the spanning states of each weight space gives the matrix representation
\begin{equation}
\func{\Delta}{L_0} = \frac{1}{3} 
\begin{pmatrix}
\lambda + \tfrac{9}{8} & -2 \\
\tfrac{1}{8} \brac{2 \lambda + 1} \brac{2 \lambda + 3} & -\lambda - \tfrac{7}{8}
\end{pmatrix}
.
\end{equation}
This matrix has eigenvalues $-\tfrac{1}{8}$ and $\tfrac{5}{24}$ for all $\lambda$.  Again, zero-grade states are forbidden from having the latter conformal dimension (\secref{secOld}), so we suspect that only the eigenstates corresponding to eigenvalue $-\tfrac{1}{8}$ are actually present in the fusion.  The other eigenstates should then be spurious states.

As in \secref{secFusLL}, we use the vanishing vectors \eqref{eqnSV1E} of $\AffOthMod{\lambda}$ in the fusion algorithm of \secref{secFus0Pre} to search for spurious states in the weight spaces.  Applying $\func{\Delta}{e_{-1}} = 0$ to $\ket{u_1} \otimes \ket{v_{m-2}}$, $\func{\Delta}{h_{-1}} = 0$ to $\ket{u_1} \otimes \ket{v_m}$ and $\func{\Delta}{f_{-1}} = 0$ to $\ket{u_1} \otimes \ket{v_{m+2}}$, we find that \eqref{eqnSV1E} leads to the spurious states
\begin{equation} \label{eqnLERel}
\ket{u_{-1}} \otimes \ket{v_{m+2}} - \frac{2 m + 3}{4} \ket{u_1} \otimes \ket{v_m} = 0 \qquad \text{(for all $m \in 2 \ZZ + \lambda$),}
\end{equation}
which must be removed from each weight space.  The weight spaces are therefore (at most) one-dimensional.

We can repeat the above exercise after replacing $\ket{u_1}$ by $\ket{u_{-1}}$, but find no further spurious states.  Similarly, the vanishing vectors \eqref{eqnSV2E} of $\AffOthMod{\lambda}$ and the vanishing singular vector \eqref{eqnL1SV} of $\AffIrrMod{1}$ (using \eqnref{eqnMaster3} for the latter)  yield nothing new, so we conclude that the relations \eqref{eqnLERel} are exhaustive.  The result of the fusion to grade $0$ is therefore an infinite-dimensional space whose $\SLA{sl}{2}$-weights belong to $2 \ZZ + \lambda + 1$ and have multiplicity one.  There is only one admissible $\AKMA{sl}{2}$-module with this zero-grade subspace, $\AffOthMod{\lambda + 1}$, so this strongly suggests that the fusion rule is
\begin{equation} \label{eqnLEFusion2}
\AffIrrMod{1} \fuse \AffOthMod{\lambda} = \AffOthMod{\lambda + 1},
\end{equation}
where the addition is of course understood modulo $2$.  We have checked that the action of the zero-modes (\eqnref{eqnMaster0}) is consistent with this conclusion.

As in \secref{secFusLL}, one is required to rule out the presence, in this fusion decomposition, of twisted modules which are not detected at grade $0$.  The argument presented there works just as well in this case, except now we use the twisted fusion rule
\begin{equation} \label{eqnL1xTwistE}
\AffIrrMod{1} \fuse \tfunc{\gamma^{-\ell}}{\AffOthMod{\lambda}} \qquad \text{($\abs{\ell} \geqslant 1$).}
\end{equation}
Termination of step \ref{Step1} follows because $\AffIrrMod{1}$ is highest weight (even relaxed highest weight would suffice).  Step \ref{Step2} again terminates essentially because it reduces to the transfer of $\SLA{sl}{2}$-weight from the infinitely many states of $\tfunc{\gamma^{-\ell}}{\AffOthMod{\lambda}}$ whose conformal dimension is minimal for their weight to the zero-grade states of $\AffIrrMod{1}$.  This can only be done twice before the latter states are annihilated.  The termination of the fusion algorithm then implies that \eqref{eqnL1xTwistE} is trivial to grade $0$, ruling out undetected twisted modules in \eqref{eqnLEFusion2}.

\eqnref{eqnLEFusion2} is therefore correct, proving our earlier claim (\secref{secOld}) that $\AffIrrMod{1}$ remains a simple current when we augment the theory by the $\AffOthMod{\lambda}$ and their spectral flow images.  Note that it follows now from associativity \cite{GabFus94} that the vacuum module $\AffIrrMod{0}$ continues to act as the fusion identity in this augmented theory.  Of course one can explicitly check this too using the grade $0$ fusion algorithm.

\subsection{Fusing $\AffOthMod{\lambda}$ and $\AffOthMod{\mu}$} \label{secFusEE}

We now turn to the fusion of $\AffOthMod{\lambda}$ with $\AffOthMod{\mu}$ (for $\lambda , \mu \in \set{0,1}$) to grade $0$.  This time, the result is found within the vector space spanned by the states
\begin{equation}
\ket{v_n} \otimes \ket{v_m} \qquad \text{($n \in 2 \ZZ + \lambda$, $m \in 2 \ZZ + \mu$).}
\end{equation}
We mention that this is infinite-dimensional and that, in contrast to the case studied in \secref{secFusLE}, the $\SLA{sl}{2}$-weight spaces are also infinite-dimensional.  We will therefore not be able to (immediately) use the action \eqref{eqnDeltaL0Grade0} of $L_0$ to analyse (heuristically) whether these states are admissible.  This also serves as a hint that perhaps the structure of this fusion product is more subtle than those we have analysed thus far.

Let us therefore repeat the analysis of \secref{secFusLE}, starting with the vanishing vectors \eqref{eqnSV1E} of $\AffOthMod{\mu}$.  Since $\func{\Delta}{J_{-1}} = 0$ identically, we can use \eqnref{eqnMaster2} to derive the relations
\begin{equation} \label{eqnEERecRels1}
\frac{\brac{2m-1} \brac{2m+3}}{16} \ket{v_{n+2}} \otimes \ket{v_{m-2}} - \frac{\brac{2m+3} n}{4} \ket{v_n} \otimes \ket{v_m} + \frac{\brac{2n-1} \brac{2n-3}}{16} \ket{v_{n-2}} \otimes \ket{v_{m+2}} = 0.
\end{equation}
We interpret these spurious states as second-order recurrence relations for the states of $\SLA{sl}{2}$-weight $m+n$.  It follows that for each weight $m+n$, these relations reduce the number of linearly independent states from infinity to just two!  We will fix these two states in each weight space by choosing $m$ (and therefore $n$) arbitrarily.  The ``basis'' states then have the form
\begin{equation}
\ket{v_n} \otimes \ket{v_m} \qquad \text{and} \qquad \ket{v_{n+2}} \otimes \ket{v_{m-2}}.
\end{equation}
Since the weight spaces are now finite-dimensional, \eqnref{eqnDeltaL0Grade0} can be applied to determine the action of $L_0$.  With respect to the basis ordering above, we find the matrix representation
\begin{equation}
\func{\Delta}{L_0} = \frac{1}{12}
\begin{pmatrix}
-\brac{2n+1} & 2m-1 \\
-\brac{2n+1} \brac{2n+3} / \brac{2m-1} & 2n+3
\end{pmatrix}
\end{equation}
which has eigenvalues $0$ and $\tfrac{1}{2}$.  This suggests that the result of fusing $\AffOthMod{\lambda}$ with $\AffOthMod{\mu}$ will involve the module $\AffIrrMod{0}$ or $\AffIrrMod{1}$.\footnote{Of course, it cannot involve both.  The $\SLA{sl}{2}$-weights of $\AffOthMod{\lambda}$ are either all even or all odd, depending on the parity of $\lambda$.  The weights of the fusion module will therefore accord with the parity of $\lambda + \mu$.}  Note however that the result is (thus far) independent of the total $\SLA{sl}{2}$ weight $m+n$, which compares poorly with the situation for $\AffIrrMod{0}$ and $\AffIrrMod{1}$.

To analyse the fusion space in more detail, we note that the relations \eqref{eqnEERecRels1} are not symmetric in $m$ and $n$.  Indeed, if we start with the vanishing vectors \eqref{eqnSV1E} of $\AffOthMod{\lambda}$ and use \eqnref{eqnMaster3}, we derive instead the slightly different relations
\begin{equation} \label{eqnEERecRels2}
\frac{\brac{2m-1} \brac{2m-3}}{16} \ket{v_{n+2}} \otimes \ket{v_{m-2}} - \frac{m \brac{2n+3}}{4} \ket{v_n} \otimes \ket{v_m} + \frac{\brac{2n-1} \brac{2n+3}}{16} \ket{v_{n-2}} \otimes \ket{v_{m+2}} = 0.
\end{equation}
Substituting \eqref{eqnEERecRels1} into \eqref{eqnEERecRels2}, we obtain
\begin{equation} \label{eqnEERecRels3}
\brac{m+n} \Bigl[ \brac{2m-1} \ket{v_{n+2}} \otimes \ket{v_{m-2}} - \brac{2n+3} \ket{v_n} \otimes \ket{v_m} \Bigr] = 0.
\end{equation}
The resulting spurious states therefore reduce the dimension of the weight spaces to $1$ except when the weight is $m+n = 0$.  In the latter case, \eqnref{eqnEERecRels3} is vacuous so the dimension remains at $2$.

It remains to study the vanishing vectors \eqref{eqnSV2E} of $\AffOthMod{\lambda}$ and $\AffOthMod{\mu}$.  These by themselves yield rather unappealing third-order recurrence relations.  However, when $m+n=0$, applying \eqref{eqnEERecRels1} reduces both these recurrence relations to the simple form
\begin{equation}
\ket{v_{n+2}} \otimes \ket{v_{m-2}} + \ket{v_n} \otimes \ket{v_m} = 0 \qquad \text{($m+n=0$).}
\end{equation}
We conclude that the weight space of weight $m+n=0$ is therefore one-dimensional.  When $m+n \neq 0$, we can apply instead \eqnref{eqnEERecRels3} to derive that
\begin{equation}
\brac{m+n-1} \brac{m+n+1} \ket{v_n} \otimes \ket{v_m} = 0 \qquad \text{($m+n \neq 0$).}
\end{equation}
We conclude from this that the weight spaces of weight $m+n = \pm 1$ are also one-dimensional, whereas those with weight not equal to $\pm 1$ (or $0$) are trivial!  Moreover, computing the action of $\func{\Delta}{L_0}$ on the remaining states with $m+n = 0$ and $\pm 1$ gives conformal dimensions $0$ and $\tfrac{1}{2}$ respectively.  As we have been unable to find any further relations, this suggests the following fusion rules:
\begin{equation} \label{eqnEEFusWrong}
\AffOthMod{\lambda} \fuse \AffOthMod{\mu} = \AffIrrMod{\lambda + \mu},
\end{equation}
where the addition is once again taken modulo $2$.  We remark that these fusion rules are consistent with associativity.

However, we recall from \secref{secOld} that $\AffOthMod{0}$ and $\AffOthMod{1}$ decompose in the $\AKMA{u}{1}$-coset theory into the \hwms{} of the triplet algebra whose \hwss{} have conformal dimensions $-\tfrac{1}{8}$ and $\tfrac{3}{8}$, respectively.  These triplet algebra modules are well known to fuse with one another to give indecomposable modules on which $L_0$ cannot be diagonalised \cite{GabRat96}.  This is responsible for the logarithmic structure of the triplet model and we would expect that this logarithmic structure is mirrored in the fusion rules of the $\AKMA{sl}{2}_{-1/2}$ theory studied here.  Thus far, we have not uncovered any trace of indecomposability --- all $\AKMA{sl}{2}$-modules considered to date are in fact irreducible.  This leads us to suspect that the fusion of $\AffOthMod{\lambda}$ and $\AffOthMod{\mu}$ is not as simple as the above grade $0$ calculation would have us believe.  We will therefore check carefully whether it is possible that this calculation might have missed contributions coming from twisted modules.

Suppose then, as in \secDref{secFusLL}{secFusLE}, that there are states associated to twisted modules appearing in the fusion of $\AffOthMod{\lambda}$ and $\AffOthMod{\mu}$ (unlike what was proposed in \eqref{eqnEEFusWrong}).  Then, we can try to detect them by studying instead the grade $0$ fusion
\begin{equation} \label{eqnExTwistE}
\AffOthMod{\lambda} \fuse \tfunc{\gamma^{-\ell}}{\AffOthMod{\mu}}
\end{equation}
with $\ell \neq 0$.  Since the twisted module above has a trivial zero-grade subspace, we may conclude, as before, that \eqnref{eqnEEFusWrong} has no twisted module corrections as long as the fusion algorithm of \secref{secFus0Pre} actually terminates for \eqref{eqnExTwistE}.

Now, step \ref{Step1} still terminates, as $\AffOthMod{\lambda}$ is a relaxed \hwm{}, but the situation for step \ref{Step2} is not so happy.  We can still reduce a state $\ket{v} \otimes \ket{w}$ so that we may assume $\ket{v}$ to be a zero-grade state of $\AffOthMod{\lambda}$ and that $\ket{w} \in \tfunc{\gamma^{-\ell}}{\AffOthMod{\mu}}$ has the minimal conformal dimension possible for its weight.  But, using step \ref{Step2} repeatedly to reduce $\ket{w}$ to certain states $\ket{w_i}$ with $i \in \ZZ_+$, as in \eqnref{eqnTheSequence}, we encounter an infinite regression.  Each iteration moves some weight from the $\ket{w_i}$ onto $\ket{v} \in \AffOthMod{\lambda}$ via the action of $J_0$ ($J = e$ or $f$), but $\ket{v}$ is never annihilated this way, so the algorithm does not terminate.  It follows that we cannot exclude the presence of unseen twisted module corrections to \eqnref{eqnEEFusWrong} by computing to grade $0$.  To study this question further, we will therefore have to bite the bullet and study fusion to higher grades, or in the more precise language of \secref{secFus0Pre}, exchange the algebra $\alg{A}^-$ for a carefully chosen subalgebra.

\section{Fusion Beyond Grade $0$} \label{secFusion}

\subsection{More General Fusion Algorithms} \label{secFusPre}

We consider the generalisation of the fusion algorithm of \secref{secFus0Pre} to higher grades.  If $\mathcal{M}$ and $\mathcal{N}$ are the modules to be fused, this means choosing a subalgebra $\alg{A}$ of $\alg{A}^-$ and determining the (vector space) quotient of the fusion module on which $\alg{A}$ acts as zero.  The general formalism of Nahm and Gaberdiel-Kausch \cite{NahQua94,GabInd96} suggests that this quotient should satisfy
\begin{equation} \label{eqnFusionAllGrades}
\frac{\mathcal{M} \fuse \mathcal{N}}{\func{\func{\Delta}{\alg{A}}}{\mathcal{M} \fuse \mathcal{N}}} \subseteq \frac{\mathcal{M}}{\alg{A}^- \mathcal{M}} \otimes \frac{\mathcal{N}}{\alg{A} \mathcal{N}}.
\end{equation}
Note that $\alg{A}^-$ still appears on the right hand side.  As before, demonstrating this inclusion amounts to showing that a certain algorithm terminates.

This fusion algorithm for $\alg{A}$ is in fact only a slight generalisation of that of \secref{secFus0Pre}.  We have now three steps which we may apply iteratively to a state $\ket{v} \otimes \ket{w} \in \mathcal{M} \otimes \mathcal{N}$, though the third is little more than an afterthought:
\begin{enumerate}
\item \label{Stp1} If $\ket{v} = J_{-n} \ket{v'}$ for some $J_{-n} \in \alg{A}^-$, then we apply \eqnref{eqnMaster3}.
\item \label{Stp2} If $\ket{w} = J_{-n_1} \cdots J_{-n_{\ell}} \ket{w'}$ for some $J_{-n_1} \cdots J_{-n_{\ell}} \in \alg{A}$, then we apply \eqnref{eqnMaster2} to expand
\begin{equation}
\func{\func{\Delta}{J_{-n_1} \cdots J_{-n_{\ell}}}}{\ket{v} \otimes \ket{w'}} = 0
\end{equation}
and substitute for $\ket{v} \otimes \ket{w}$.
\item \label{Stp3} It may happen that $\ket{w} = J_{-n_1} \cdots J_{-n_{\ell}} \ket{w'} \in \alg{A} \mathcal{N}$ (formally), but $J_{-n_1} \cdots J_{-n_{\ell}} \notin \alg{A}$.  This occurs when $J_{-n_1} \cdots J_{-n_{\ell}} \ket{w'}$ is one term of a vanishing state of $\mathcal{N}$, the other terms of which belong to $\alg{A} \mathcal{N}$.  Then, we simply use this vanishing to substitute for $J_{-n_1} \cdots J_{-n_{\ell}} \ket{w'}$.
\end{enumerate}
We will see an example in the next subsection of step \ref{Stp3} being used.  As in \secref{secFus0Pre}, it is not hard to verify that this algorithm is guaranteed to terminate when $\mathcal{M}$ and $\mathcal{N}$ are (relaxed) \hwms{}.  In general however, termination is a very subtle affair as we saw in \secref{secFusEE}.

\subsection{$\AffOthMod{\lambda} \fuse \AffOthMod{\mu}$ Revisited} \label{secFusEEAgain}

Our first task, and it is a very important one, is to choose the algebra $\alg{A}$ that appears in the fusion algorithm.  One might expect, especially if one is familiar with similar fusion computations for the Virasoro algebra, that a natural choice would be the subalgebra generated by all products of negative modes whose indices sum to $-2$ or less (this would be a good candidate for fusion to grade $1$).  However, computations with this $\alg{A}$ tell us little more than the grade $0$ computations of \secref{secFusion0}, and the reason is because this $\alg{A}$ is likewise blind to the appearance of non-trivial twisted modules in the fusion results.

We will therefore first consider taking the algebra $\alg{A}$ to be that generated by $e_{-n-1}$, $h_{-n}$ and $f_{-n}$ with $n \geqslant 1$.  We will refer to the corresponding fusion algorithm as fusing to grade $\left(1,0,0\right)$.  Since $e_{-1}^j \notin \alg{A}$ for any $j$, we expect that fusing in this way will expose twisted modules of the form $\tfunc{\gamma^2}{\AffIrrMod{\mu}}$ and $\tfunc{\gamma}{\AffOthMod{\mu}}$.  Other twisted modules might not, however, be visible with this approach.

Having chosen $\alg{A}$, we now compute.  As in \secDref{secFusLE}{secFusEE}, $\AffOthMod{\lambda} / \bigl( \alg{A}^- \AffOthMod{\lambda} \bigr)$ consists of just the zero-grade states $\ket{v_n}$ of $\AffOthMod{\lambda}$.  The quotient $\AffOthMod{\mu} / \bigl( \alg{A} \AffOthMod{\mu} \bigr)$ should consist of the zero-grade states $\ket{v_m}$ of $\AffOthMod{\mu}$ and their $e_{-1}$-descendants.  However, \eqnref{eqnSV1E} lets us write
\begin{equation} \label{eqnRemoveE}
e_{-1} \ket{v_{m-2}} = \frac{4}{2 m - 1} h_{-1} \ket{v_m} - \frac{16}{\brac{2 m - 1} \brac{2 m + 3}} f_{-1} \ket{v_{m + 2}} \in \alg{A} \AffOthMod{\mu},
\end{equation}
for all $m$.  It follows that $\AffOthMod{\mu} / \bigl( \alg{A} \AffOthMod{\mu} \bigr)$ likewise consists of just the zero-grade states $\ket{v_m}$ of $\AffOthMod{\mu}$.

\eqnref{eqnFusionAllGrades} now tells us that the fusion quotient is contained within the vector space spanned by the
\begin{equation}
\ket{v_n} \otimes \ket{v_m} \qquad \text{($n \in 2 \ZZ + \lambda$, $m \in 2 \ZZ + \mu$).}
\end{equation}
Before beginning the computations, we pause to consider what the result will be if the suggested fusion rules \eqref{eqnEEFusWrong} are correct.  We therefore illustrate the quotients $\AffIrrMod{\lambda + \mu} / \bigl( \alg{A} \AffIrrMod{\lambda + \mu} \bigr)$ for $\lambda + \mu = 0,1$ in \figref{figLGrade100}, along with the actions of $e_{-1}$, $e_0$, $f_0$ and $f_1$.  There are several comments in order here.  First, \eqnref{eqnL1SV} gives
\begin{equation}
4 e_{-1}^2 \ket{u_{-1}} = \brac{2 h_{-1} e_{-1} - 7 e_{-2}} \ket{u_1} \in \alg{A} \AffIrrMod{1},
\end{equation}
explaining why the picture for $\AffIrrMod{1}$ has only one infinite string of states, $e_{-1}^j \ket{u_1}$, rather than two.  Second, $e_0$ and $f_1$ do not preserve $\alg{A}$ (under the adjoint action), hence they are not truly well-defined on the quotients under consideration.  In particular, in any quotient of the form $\mathcal{N} / \bigl( \alg{A} \mathcal{N} \bigr)$, we may write
\begin{equation}
e_0 \ket{v} = e_0 \bigl( \ket{v} + h_{-1} \ket{w} \bigr) = e_0 \ket{v} + h_{-1} e_0 \ket{w} - 2 e_{-1} \ket{w} = e_0 \ket{v} - 2 e_{-1} \ket{w}
\end{equation}
for any $\ket{w}$.  Assuming (without any loss of generality) that $\ket{v}$ has a definite $\SLA{sl}{2}$-weight and conformal dimension, it follows that we can only define $e_0 \ket{v}$ unambiguously if the $\AKMA{sl}{2}$-weight space of $e_0 \ket{v}$ has trivial intersection with the image of $e_{-1}$.  Similarly, $f_1$ is only defined on states $\ket{v}$ for which the $\AKMA{sl}{2}$-weight space of $f_1 \ket{v}$ has trivial intersection with the image of $f_0$.  It follows that both $e_0$ and $f_1$ can be defined on all of $\AffIrrMod{0} / \bigl( \alg{A} \AffIrrMod{0} \bigr)$ and $\AffIrrMod{1} / \bigl( \alg{A} \AffIrrMod{1} \bigr)$ except for the state which corresponds to the south-east corner of the parallelogram in \figref{figLGrade100}.  Note however that $e_{-1}$, $h_0$, $f_0$ and $L_0$ are perfectly well-defined.

{
\psfrag{L0}[][]{$\AffIrrMod{0} / \bigl( \alg{A} \AffIrrMod{0} \bigr)$}
\psfrag{L1}[][]{$\AffIrrMod{1} / \bigl( \alg{A} \AffIrrMod{1} \bigr)$}
\begin{figure}
\begin{center}
\includegraphics[width=8cm]{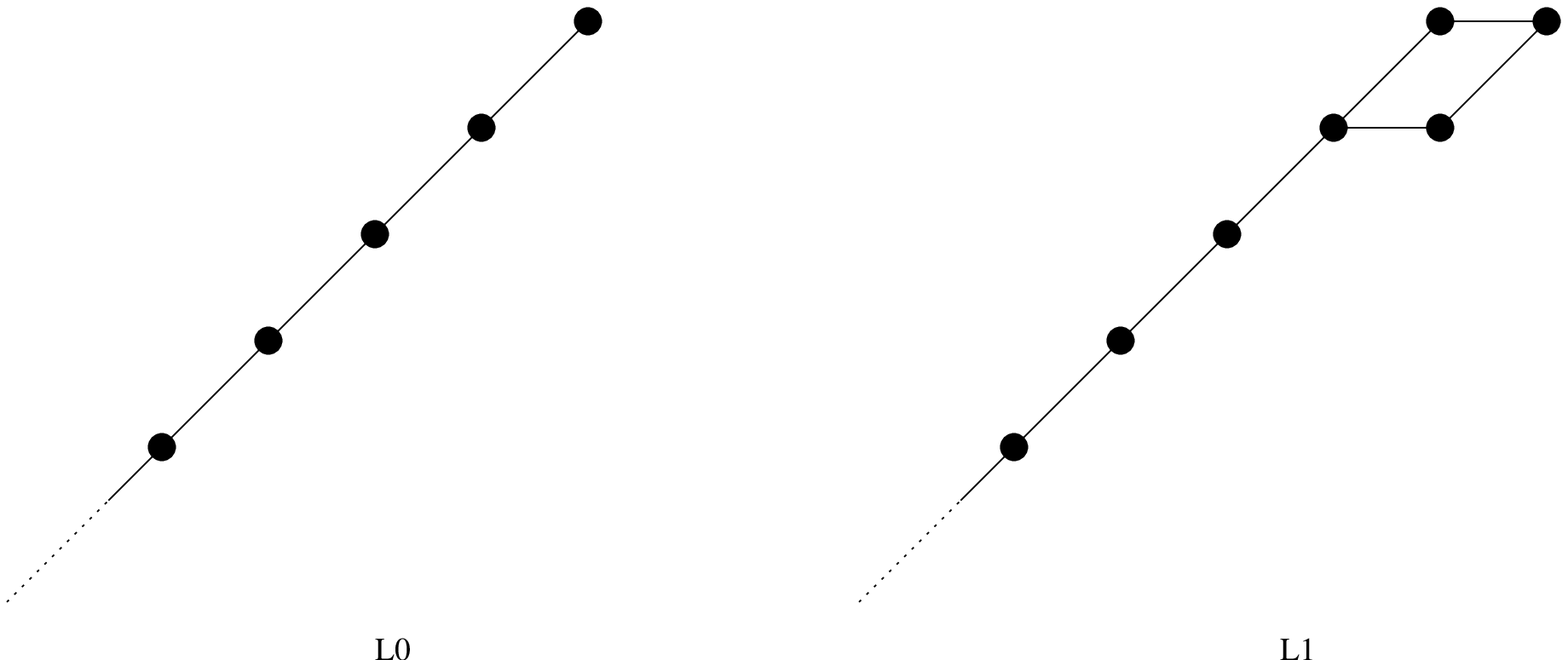}
\caption{Schematic illustrations of the states of the (vector space) quotients $\AffIrrMod{0} / \bigl( \alg{A} \AffIrrMod{0} \bigr)$ and $\AffIrrMod{1} / \bigl( \alg{A} \AffIrrMod{1} \bigr)$.  As usual, the conformal dimension increases from top to bottom and the $\SLA{sl}{2}$-weight increases from right to left.  Each dot represents a state for the corresponding weight space (the multiplicity of which is always one).  The lines represent the actions of $e_{-1}$ (south-west), $e_0$ (west), $f_0$ (east) and $f_1$ (north-east).} \label{figLGrade100}
\end{center}
\end{figure}
}

We now begin the computations.  As in \secref{secFusEE}, the weight spaces are infinite-dimensional.  One difference is that we cannot start by using the vanishing states \eqref{eqnSV1E} of $\AffOthMod{\mu}$.  These vectors have already been used to reduce the states of $\AffOthMod{\mu} / \bigl( \alg{A} \AffOthMod{\mu} \bigr)$ to the zero-grade states of $\AffOthMod{\mu}$.  One can check that trying to use them further takes us around in circles.  Instead, we must start with the corresponding vanishing states of $\AffOthMod{\lambda}$.  Applying \eqnref{eqnMaster3} to the tensor product of such a state with $\ket{v_m}$ and using the fusion algorithm of \secref{secFusPre}, we arrive at a third-order recurrence relation for each weight space.  The weight spaces are thereby reduced to having dimension $3$.

We continue our search for spurious states, now using the vanishing states \eqref{eqnSV2E} of $\AffOthMod{\mu}$.  These turn out to yield an independent set of spurious states which reduce the dimension of each weight space by one.  Furthermore, applying $\func{\Delta}{f_0}$ (which is well-defined) to these spurious states yields new ones except when the weight is $m+n = \pm 1$.  We have found no further spurious states using these vanishing states, nor by exploiting those of $\AffOthMod{\lambda}$, so we conclude that the weight spaces of the fusion product are one-dimensional unless the weight is $\pm 1$, in which case they have dimension $2$.

We can now determine the action of $L_0$ on the weight spaces using \eqnref{eqnMaster4}.  Applying the Sugawara form of $L_{-1}$ and \eqnref{eqnMaster3} again, we find that
\begin{align} \label{eqnL0ActionAgain}
\func{\Delta}{L_0} &= L_0 \otimes \wun + \wun \otimes L_0 + \frac{1}{3} \brac{h_{-1} h_0 - 2 e_{-1} f_0 - 2 f_{-1} e_0} \otimes \wun \notag \\
&= L_0 \otimes \wun + \wun \otimes L_0 + \frac{1}{3} \brac{h_0 \otimes h_0 - 2 e_0 \otimes f_0 - 2 f_0 \otimes e_0 - 2 e_0 f_0 \otimes \wun - 2 f_0 \otimes e_{-1}}.
\end{align}
We should be concerned that $e_0$ may not be well-defined in the second factor where it acts upon $\AffOthMod{\mu} / \bigl( \alg{A} \AffOthMod{\mu} \bigr)$.  As remarked above however, this is only a problem when the result of applying $e_0$ has the same weight and conformal dimension as a non-trivial element of the image of $e_{-1}$.  It is not hard to check that \eqnref{eqnRemoveE} implies that this image has already been set to zero, so $e_0$ is in fact well-defined on $\AffOthMod{\mu} / \bigl( \alg{A} \AffOthMod{\mu} \bigr)$.  It follows that the above action of $L_0$ makes sense.

Computing the action of $L_0$ on a generic weight space now gives a conformal dimension of $\tfrac{1}{2} \brac{m+n}$, where $m+n \neq \pm 1$ is the corresponding (generic) weight.  For $m+n = 1$, $\func{\Delta}{L_0}$ is a matrix with eigenvalues $\tfrac{1}{2}$ and $\tfrac{3}{2}$, and for $m+n = -1$ the eigenvalues are $\tfrac{1}{2}$ and $\tfrac{-1}{2}$.  The fact that these states have conformal dimensions which are unbounded above and below proves that \eqnref{eqnEEFusWrong} is incorrect --- twisted modules do contribute to the fusion.  We illustrate the weight spaces of the fusion module to grade $\left(1,0,0\right)$ in \figref{figEEGrade100}.  There, as in \figref{figLGrade100}, we indicate the action of $e_{-1}$, $e_0$, $f_0$ and $f_1$.  Note that $e_0$ and $f_1$ are well-defined except on the state of $\SLA{sl}{2}$-weight $1$ and conformal dimension $\tfrac{3}{2}$ (the south-east ``corner'' of the parallelogram of states appearing in \figref{figEEGrade100}).

{
\psfrag{L0}[][]{$\lambda + \mu = 0$}
\psfrag{L1}[][]{$\lambda + \mu = 1$}
\begin{figure}
\begin{center}
\includegraphics[width=10cm]{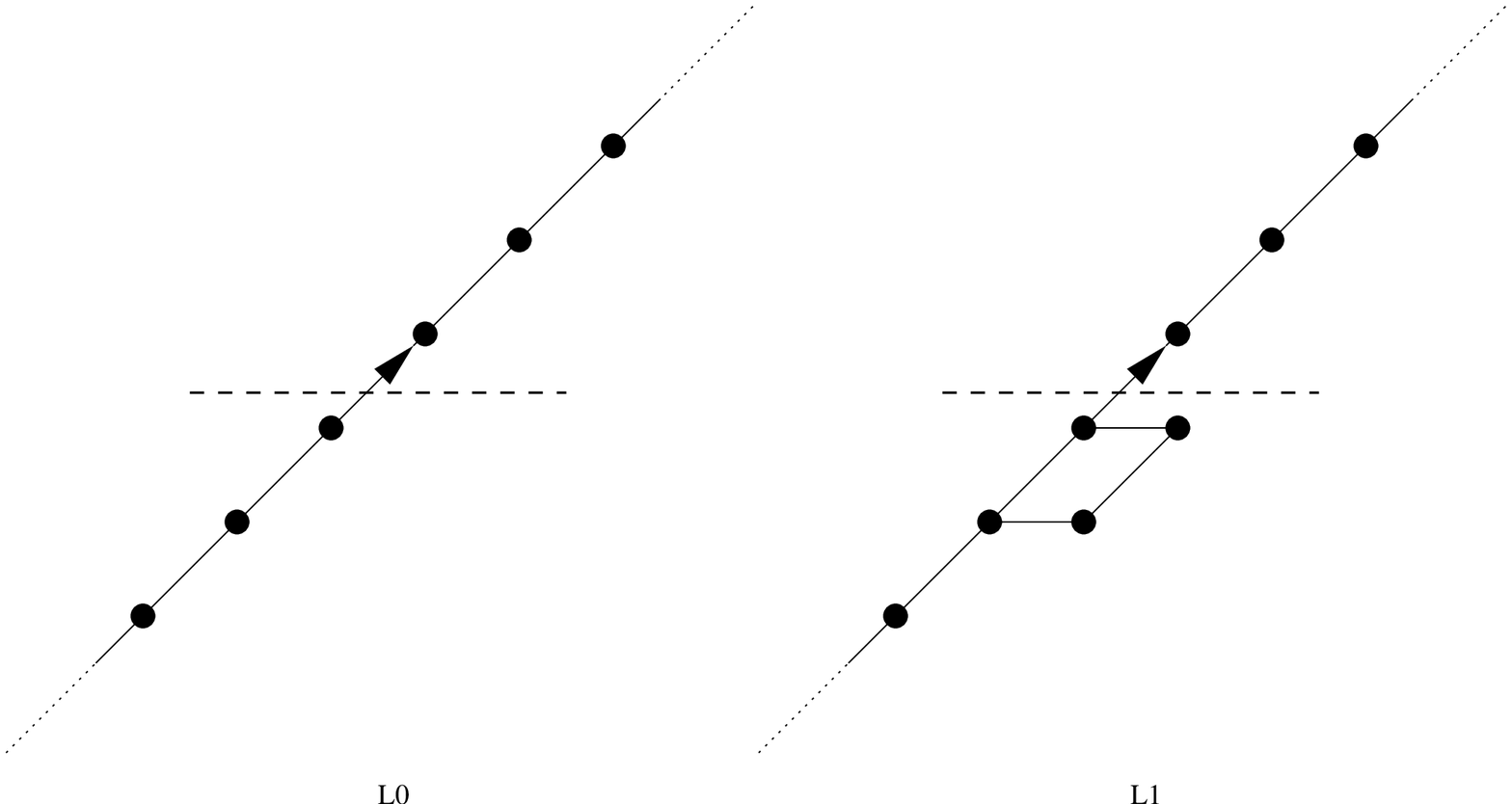}
\caption{Schematic illustrations of the states of the (vector space) quotients of the fusion module $\AffOthMod{\lambda} \fuse \AffOthMod{\mu}$ to grade $\left(1,0,0\right)$ for $\lambda + \mu = 0$ and $\lambda + \mu = 1$ (modulo $2$).  The conventions are as in \figref{figLGrade100} and the north-east arrow signifies that $e_{-1}$ is acting trivially whereas $f_1$ is not.  The dashed lines serve to delineate the point below which the result resembles that of \figref{figLGrade100}.} \label{figEEGrade100}
\end{center}
\end{figure}
}

Now, if we twist these results by the spectral flow automorphism $\gamma^{-1}$, we end up with two weight space configurations whose conformal dimensions are uniformly bounded below (by $-\tfrac{1}{8}$) and whose $\SLA{sl}{2}$-weights are half-integers.  Indeed, when $\lambda + \mu = 0$, the configuration suggests that the twisted result is indecomposable with a submodule isomorphic to $\tfunc{\gamma}{\AffIrrMod{1}} = \AffIrrMod{-3/2}$ and whose quotient by this submodule is isomorphic to $\tfunc{\gamma^{-1}}{\AffIrrMod{0}} = \AffIrrMod{-1/2}^*$ (the ``$*$'' indicates the conjugate representation).  This indecomposable module therefore has a zero-grade subspace with non-trivial weight spaces of arbitrarily large weights, both positive and negative.  In this, it resembles the modules $\AffOthMod{\lambda}$, except that its zero-grade subspace contains an $\SLA{sl}{2}$-\hws{} of weight $-\tfrac{3}{2}$.  From \secref{secOld}, we therefore identify this indecomposable as $\AffOthMod{-3/2}^+$ (recall that the ``$+$'' indicates that there is a \hws{} of the given weight $-\tfrac{3}{2}$).  When \mbox{$\lambda + \mu = 1$}, the corresponding twisted result is obtained by swapping $\AffIrrMod{0}$ and $\AffIrrMod{1}$.  This results in the indecomposable module $\AffOthMod{-1/2}^+$ whose zero-grade subspace has an $\SLA{sl}{2}$-\hws{} of weight $-\tfrac{1}{2}$.

Undoing the spectral flow, we see that the grade $\left(1,0,0\right)$ result suggests that the true fusion rules take the form
\begin{equation} \label{eqnEEFusStillWrong}
\AffOthMod{\lambda} \fuse \AffOthMod{\mu} = 
\begin{cases}
\tfunc{\gamma}{\AffOthMod{-3/2}^+} & \text{if $\lambda + \mu = 0 \pmod{2}$,} \\
\tfunc{\gamma}{\AffOthMod{-1/2}^+} & \text{if $\lambda + \mu = 1 \pmod{2}$.}
\end{cases}
\end{equation}
This result strongly confirms our suspicion that the result of fusing $\AffOthMod{\lambda}$ with $\AffOthMod{\mu}$ involves a reducible yet indecomposable module.  However, the above identification of the fusion cannot be correct either!  Applying \eqnref{eqnFusionAssumption} to the left hand side of \eqnref{eqnEEFusStillWrong}, we discover that the right hand side should be self-conjugate.  What we have concluded above is not.  This is because we have based our conclusion on the result of fusing to grade $\left(1,0,0\right)$, corresponding to an algebra $\alg{A}$ which is not invariant under the conjugation automorphism $\mathsf{w}$.  Indeed, if we had fused to grade $\left(0,0,1\right)$ instead, corresponding to $\alg{A}$ being generated by $e_{-n}$, $h_{-n}$ and $f_{-n-1}$ with $n \geqslant 1$, then we would be drawing mirror images of the weight space configurations of \figDref{figLGrade100}{figEEGrade100}.  We conclude that we are yet to unravel the full structure of the fusion module.

We therefore generalise the above fusion computations to grade $\left(1,0,1\right)$, that is we redefine $\alg{A}$ to be generated by the (conjugation-invariant set) $e_{-n-1}$, $h_{-n}$ and $f_{-n-1}$ with $n \geqslant 1$.  It follows that the fusion quotient that we will uncover will be contained within the vector space spanned by the states
\begin{equation}
\ket{v_n} \otimes \ket{v_m}, \qquad \text{and} \qquad \ket{v_n} \otimes e_{-1} \ket{v_{m-2}} \qquad \text{($n \in 2 \ZZ + \lambda$, $m \in 2 \ZZ + \mu$).}
\end{equation}
Here, we have used the vanishing states \eqref{eqnSV1E} and \eqref{eqnSV2E} of $\AffOthMod{\mu}$ to eliminate states of the form $f_{-1} \ket{v_m}$ and $e_{-1}^2 \ket{v_m}$ respectively.  We will therefore not be able to use these vanishing states to construct spurious states.

We instead apply \eqnref{eqnMaster3} to the tensor product of a vanishing state \eqref{eqnSV1E} of $\AffOthMod{\lambda}$ and the state $\ket{v_m} \in \AffOthMod{\mu}$.  As before, this yields non-trivial spurious states in each weight space.  Repeating this, with $\ket{v_m}$ replaced by $e_{-1} \ket{v_{m-2}}$, yields independent spurious states which, together with the first set, define recurrences that reduce the dimension of each weight space to six.  Now, the actions of $e_0$ and $f_0$ are not well-defined on the fusion quotient, but those of $e_{-1}$ and $f_{-1}$ are.  This turns out to be very useful --- applying $\func{\Delta}{e_{-1}}$ repeatedly to the spurious states which have already been determined yields a complete set of spurious states.  No further constraints have been found, not even if we use the vanishing states \eqref{eqnSV2E} of $\AffOthMod{\lambda}$.

The resulting weight spaces are found to be generically two-dimensional, whereas those of weight $0$, $\pm1$ or $\pm2$ have dimension $3$.  Computing $L_0$ with the appropriate generalisation of \eqnref{eqnL0ActionAgain}, we find that the generic weight space corresponding to $\SLA{sl}{2}$-weight $m+n$ is spanned by $L_0$-eigenstates of conformal dimensions $\pm \tfrac{1}{2} \brac{m+n}$.  The spaces of weight $\pm 2$ yield eigenstates of dimension $-1$, $1$ and $3$, and weights $\pm 1$ give dimensions $-\tfrac{1}{2}$, $\tfrac{1}{2}$ and $\tfrac{3}{2}$.  The most interesting weight space is, however, that of weight $0$.  Here, the computations reveal two $L_0$-eigenstates of dimensions $0$ and $2$ and one generalised eigenstate of dimension $0$.  Thus, $L_0$ is not diagonalisable on this weight space, possessing instead a Jordan cell of rank $2$.  We illustrate the fusion module modulo the action of $\alg{A}$ in \figref{figEEGrade101}, along with the observed (well-defined) actions of $e_{-1}$ and $f_{-1}$.

{
\psfrag{L0}[][]{$\lambda + \mu = 0$}
\psfrag{L1}[][]{$\lambda + \mu = 1$}
\psfrag{x}[][]{$\ket{\omega_0}$}
\psfrag{y}[][]{$\ket{y_0}$}
\psfrag{x+}[][]{$\ket{x^+_0}$}
\psfrag{x-}[][]{$\ket{x^-_0}$}
\begin{figure}
\begin{center}
\includegraphics[width=14cm]{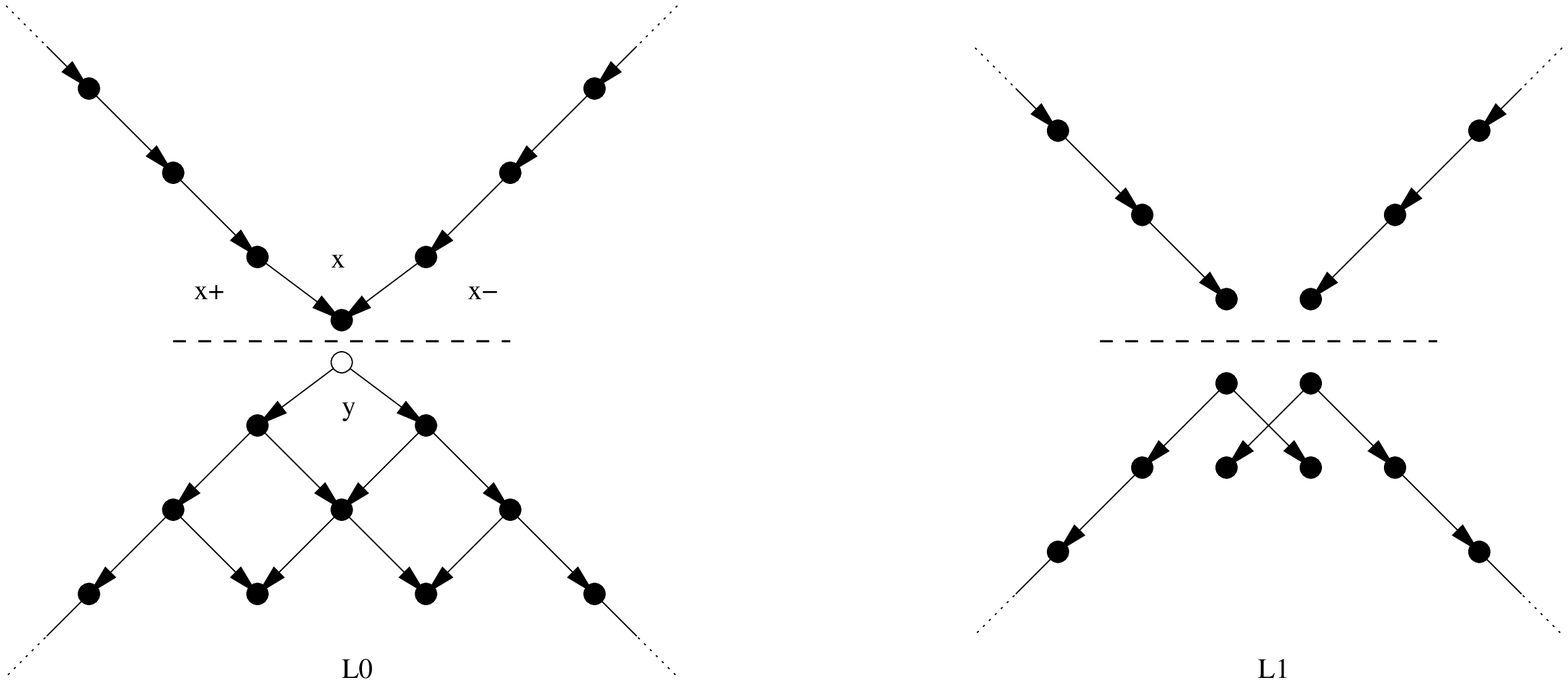}
\caption{Schematic illustrations of the states of the (vector space) quotients of the fusion module $\AffOthMod{\lambda} \fuse \AffOthMod{\mu}$ to grade $\left(1,0,1\right)$ for $\lambda + \mu = 0$ and $\lambda + \mu = 1$ (modulo $2$).  The conventions are as in \figDref{figLGrade100}{figEEGrade100}, except that we only indicate the actions of $e_{-1}$ and $f_{-1}$ (hence the arrows).  The dashed lines serve to delineate the point below which we can identify $\AffIrrMod{\lambda + \mu}$.  We also indicate for convenience the splitting of the weight-zero, dimension-zero space into the $L_0$-eigenvector $\ket{\omega_0}$ and its Jordan partner $\ket{y_0}$ (in white), as well as the states $\ket{x^{\pm}_0}$.} \label{figEEGrade101}
\end{center}
\end{figure}
}

\subsection{Analysis when $\lambda + \mu = 0$} \label{secAnalysis1}

Let us restrict ourselves to the analysis of the above fusion computation in the case when $\lambda + \mu = 0$ modulo $2$.  We will come back to the case $\lambda + \mu = 1$ later, as the observed lack of Jordan cells for this fusion quotient suggests that we still have further structure to uncover.  Consider therefore the state $\ket{x^+_0}$ appearing at grade $\left(1,0,1\right)$ whose $\SLA{sl}{2}$-weight is $2$ and whose conformal dimension is $-1$.  We suppose that this state has norm $1$.  Applying $f_{-1}$, the observed result is the weight $0$, dimension $0$ eigenstate $\ket{\omega_0}$ of $L_0$.  We normalise this eigenstate by defining
\begin{equation}
\ket{\omega_0} = f_{-1} \ket{x^+_0}.
\end{equation}
The above fusion computations indicate that $f_{-1} \ket{\omega_0} = 0$ (to grade $\left(1,0,1\right)$).  Similarly, we introduce the state $\ket{x^-_0}$ of weight $-2$ and dimension $-1$ by requiring that $\ket{\omega_0} = e_{-1} \ket{x^-_0}$.  As we know that the corresponding weight space is one-dimensional, this defines $\ket{x^-_0}$ uniquely.  We mention that $e_{-1} \ket{\omega_0} = 0$ (to grade $\left(1,0,1\right)$) as well.  Finally, we denote the Jordan partner of $\ket{\omega_0}$ by $\ket{y_0}$.  It satisfies
\begin{equation}
L_0 \ket{y_0} = \ket{\omega_0},
\end{equation}
which serves as a normalisation, though this relation does not define $\ket{y_0}$ uniquely.  Within this (generalised) weight space of weight $0$ and dimension $0$, we are free to add arbitrary multiples of $\ket{\omega_0}$ to $\ket{y_0}$ without affecting the latter's defining property.

We first consider the (induced) action of the conjugation automorphism $\mathsf{w}$ defined in \eqnref{eqnAuts}.  As the fusion module we are studying is self-conjugate, we must choose how to identify the module with its image under $\mathsf{w}$.  This will be achieved by setting
\begin{equation} \label{eqnS0Conj}
\tfunc{\mathsf{w}}{\ket{y_0}} = \ket{y_0}.
\end{equation}
The other states defined above will therefore satisfy
\begin{equation}
\tfunc{\mathsf{w}}{\ket{\omega_0}} = \ket{\omega_0} \qquad \text{and} \qquad \tfunc{\mathsf{w}}{\ket{x_0^{\pm}}} = \ket{x_0^{\mp}}.
\end{equation}
Indeed, $\ket{y_0}$ will turn out to generate the fusion module, so \eqnref{eqnS0Conj} completely defines the action of $\mathsf{w}$.\footnote{We mention that one could have tried to instead define the action of $\mathsf{w}$ by $\tfunc{\mathsf{w}}{\ket{y_0}} = \ket{\omega_0}$.  This is not correct, as the self-conjugacy of $L_0$ now implies that $\tfunc{\mathsf{w}}{\ket{\omega_0}} = 0$, contradicting $\mathsf{w}^2$ being the identity.  However, such an action is intimately related to the \emph{contragredient} dual of the fusion module.  This highlights nicely the fact that conjugate and contragredient need not coincide in a \cft{}.}

Consider now the effect of applying the spectral flow automorphism $\gamma$ to these states.  In particular, $\tfunc{\gamma}{\ket{x^+_0}}$ has weight $\tfrac{3}{2}$ and dimension $-\tfrac{1}{8}$ whereas $\tfunc{\gamma}{\ket{\omega_0}}$ has weight $-\tfrac{1}{2}$ and dimension $-\tfrac{1}{8}$.  They are related by the action of $f_0$.  It now follows from the classification of admissible relaxed \hwms{} (\secref{secOld}) that $\tfunc{\gamma}{\ket{\omega_0}}$ must be a \hws{} generating an irreducible module isomorphic to $\AffIrrMod{-1/2} = \tfunc{\gamma}{\AffIrrMod{0}}$.  Quotienting by the submodule generated by $\tfunc{\gamma}{\ket{\omega_0}}$, we find that the equivalence class of $\tfunc{\gamma}{\ket{x^+_0}}$ must generate an irreducible module isomorphic to $\AffIrrMod{-3/2}^* = \tfunc{\gamma^{-1}}{\AffIrrMod{1}}$ (refer to \figref{figSpecFlow}).  Summarising, the state $\tfunc{\gamma}{\ket{x^+_0}}$ generates an indecomposable submodule isomorphic to $\AffOthMod{-1/2}^+$ of the $\gamma$-twisted fusion module.  This indecomposable is described by the short exact sequence
\begin{equation}
\dses{\tfunc{\gamma}{\AffIrrMod{0}}}{\AffOthMod{-1/2}^+}{\tfunc{\gamma^{-1}}{\AffIrrMod{1}}}.
\end{equation}

Undoing the spectral flow, we have therefore deduced from the fusion results to grade $\left(1,0,1\right)$ that $\ket{\omega_0}$ generates a submodule isomorphic to $\AffIrrMod{0}$ and $\ket{x^+_0}$ generates an indecomposable submodule $\tfunc{\gamma^{-1}}{\AffOthMod{-1/2}^+}$ which is described by the short exact sequence
\begin{equation}
\dses{\AffIrrMod{0}}{\tfunc{\gamma^{-1}}{\AffOthMod{-1/2}^+}}{\tfunc{\gamma^{-2}}{\AffIrrMod{1}}}.
\end{equation}
We should therefore identify $\ket{\omega_0}$ with the vacuum $\ket{0}$ (since the vacuum is supposed to be unique).  A similar argument demonstrates that $\ket{x^-_0}$ likewise generates an indecomposable submodule which is isomorphic to $\tfunc{\gamma}{\AffOthMod{1/2}^-}$.  The corresponding short exact sequence is
\begin{equation}
\dses{\AffIrrMod{0}}{\tfunc{\gamma}{\AffOthMod{1/2}^-}}{\tfunc{\gamma^2}{\AffIrrMod{1}}}.
\end{equation}
Note that the two indecomposables $\tfunc{\gamma^{-1}}{\AffOthMod{-1/2}^+}$ and $\tfunc{\gamma}{\AffOthMod{1/2}^-}$ are manifestly conjugate to one another.

One can ask how we were able to conclude that $\ket{\omega_0}$ generates a submodule isomorphic to $\AffIrrMod{0}$ when the above fusion calculations show that $e_{-1} \ket{\omega_0} = f_{-1} \ket{\omega_0} = 0$.  The resolution is that the states $e_{-1} \ket{\omega_0}$ and $f_{-1} \ket{\omega_0}$, and in fact all other states descended from $\ket{\omega_0}$, are in the image of the grade-$\left(1,0,1\right)$ algebra $\alg{A}$, hence are set to zero in the fusion quotient that we have computed above.  This can be demonstrated explicitly:  Since $f_{-2}^2 \ket{x^+_0} = 0$ in $\tfunc{\gamma^{-2}}{\AffIrrMod{1}}$, it follows that $e_1^2 f_{-2}^2 \ket{x^+_0} = 0$ in $\tfunc{\gamma^{-2}}{\AffIrrMod{1}}$ and so
\begin{equation}
e_1^2 f_{-2}^2 \ket{x^+_0} = \alpha e_{-1} \ket{\omega_0} \qquad \text{in $\tfunc{\gamma^{-1}}{\AffOthMod{-1/2}^+}$, for some $\alpha$,}
\end{equation}
by weight space considerations.  In fact, applying $f_1$ to both sides yields $\alpha = 8$.  Commuting the $e_1$ modes to the right now gives
\begin{equation}
8 e_{-1} \ket{\omega_0} = e_1^2 f_{-2}^2 \ket{x^+_0} = \brac{f_{-2} e_1^2 f_{-2} - 2 h_{-1} e_1 f_{-2} + 2 f_{-2} e_0 - 2 h_{-2}} \ket{x^+_0},
\end{equation}
which clearly vanishes to grade $\left(1,0,1\right)$.  The corresponding conclusion for $f_{-1} \ket{\omega_0}$ follows similarly.

It remains to consider the Jordan partner state $\ket{y_0}$.  From \figref{figEEGrade101}, this weight $0$, dimension $0$ state appears to generate a \hwm{} much like $\AffIrrMod{0}$.  More precisely, if we quotient the fusion module by the submodule generated by $\ket{x^+_0}$ and $\ket{x^-_0}$, then what remains should be isomorphic to $\AffIrrMod{0}$.  That the grade $4$ singular vector does indeed vanish in this quotient is deducible from its explicit form \eqref{eqnSVL0} and \figref{figEEGrade101} --- if this singular vector did not vanish then we would observe dimension $4$ states of weights $0$, $\pm 2$ and $\pm 4$ when computing to grade $\left(1,0,1\right)$.

We therefore finally identify the result of fusing the modules $\AffOthMod{\lambda}$ and $\AffOthMod{\mu}$ when $\lambda + \mu = 0$.  We will write the result in the form
\begin{equation} \label{eqnEEFus0}
\AffOthMod{\lambda} \fuse \AffOthMod{\mu} = \AffStagMod{0} \qquad \text{if $\lambda + \mu = 0 \pmod{2}$,}
\end{equation}
where $\AffStagMod{0}$ is an indecomposable module with two composition series,
\begin{subequations}
\begin{gather}
\phantom{\text{and}} \qquad 0 \subset \AffIrrMod{0} \subset \tfunc{\gamma^{-1}}{\AffOthMod{-1/2}^+} \subset \tfunc{\gamma^{-1}}{\AffOthMod{-1/2}^+} + \tfunc{\gamma}{\AffOthMod{1/2}^-} \subset \AffStagMod{0} \\
\text{and} \qquad 0 \subset \AffIrrMod{0} \subset \tfunc{\gamma}{\AffOthMod{1/2}^-} \subset \tfunc{\gamma^{-1}}{\AffOthMod{-1/2}^+} + \tfunc{\gamma}{\AffOthMod{1/2}^-} \subset \AffStagMod{0},
\end{gather}
\end{subequations}
which are related by conjugation.  The composition factors --- these are the quotients of the successive submodules of a composition series --- are the (in order) irreducible modules
\begin{equation}
\AffIrrMod{0}, \ \tfunc{\gamma^{-2}}{\AffIrrMod{1}}, \ \tfunc{\gamma^2}{\AffIrrMod{1}}, \ \AffIrrMod{0} \qquad \text{and} \qquad \AffIrrMod{0}, \ \tfunc{\gamma^2}{\AffIrrMod{1}}, \ \tfunc{\gamma^{-2}}{\AffIrrMod{1}}, \ \AffIrrMod{0},
\end{equation}
respectively.  Alternatively, we can describe $\AffStagMod{0}$ in terms of a short exact sequence involving two (twisted, relaxed) \hwms{}.  Specifically, there are two such short exact sequences,
\begin{subequations}
\begin{gather} \label{eqnSESS0}
\phantom{\text{and}} \qquad \dses{\tfunc{\gamma^{-1}}{\AffOthMod{-1/2}^+}}{\AffStagMod{0}}{\tfunc{\gamma}{\AffOthMod{-3/2}^+}} \\
\text{and} \qquad \dses{\tfunc{\gamma}{\AffOthMod{1/2}^-}}{\AffStagMod{0}}{\tfunc{\gamma^{-1}}{\AffOthMod{3/2}^-}},
\end{gather}
\end{subequations}
again related by conjugation.  By analogy with similar indecomposable modules for the Virasoro algebra \cite{RohRed96,RidSta09}, we will therefore refer to the module $\AffStagMod{0}$ as a \emph{staggered module}.  We illustrate $\AffStagMod{0}$ with some of its multiplicities in \figref{figS0}.

{
\psfrag{L0}[][]{$\AffIrrMod{0}$}
\psfrag{S0}[][]{$\AffStagMod{0}$}
\psfrag{E+}[][]{$\tfunc{\gamma^{-2}}{\AffIrrMod{1}}$}
\psfrag{E-}[][]{$\tfunc{\gamma^2}{\AffIrrMod{1}}$}
\setlength{\extrarowheight}{6mm}
\begin{figure}
\begin{center}
\parbox{6.5cm}{
\includegraphics[width=6cm]{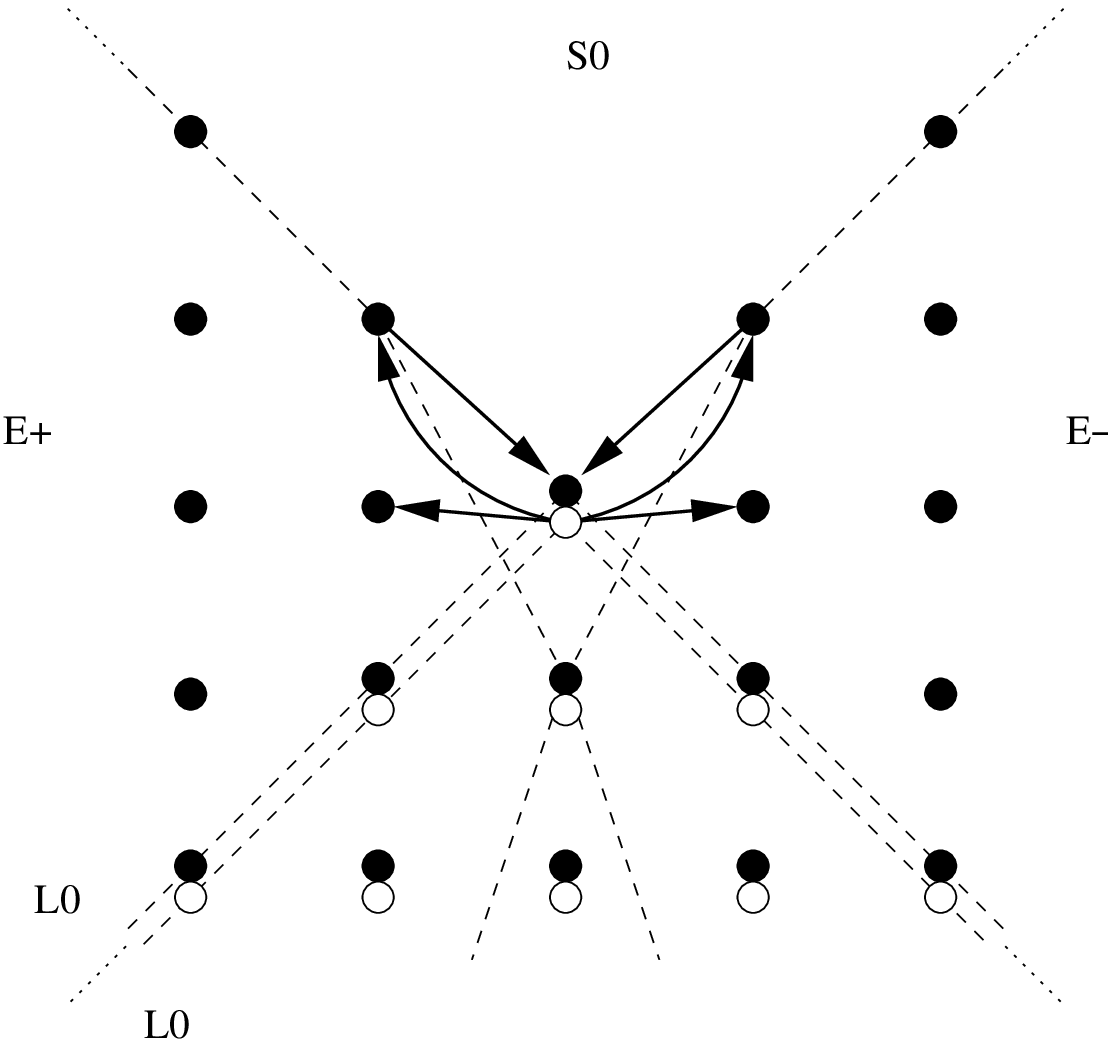}
}
\hspace{10mm}
\parbox{6.8cm}{
\begin{tabular}{*{5}{@{}C@{\hspace{2mm}}}}
\brac{1,0,0} & & & & \brac{0,0,1} \\
\brac{2,0,0} & \brac{1,0,0} & & \brac{0,0,1} & \brac{0,0,2} \\
\brac{5,0,0} & \brac{2,0,0} & \brac{0,2,0} & \brac{0,0,2} & \brac{0,0,5} \\
\brac{10,0,0} & \brac{4,2,0} & \brac{1,2,1} & \brac{0,2,4} & \brac{0,0,10} \\
\brac{19,2,0} & \brac{8,4,0} & \brac{2,6,2} & \brac{0,4,8} & \brac{0,2,19}
\end{tabular}
}
\caption{A schematic illustration of the states of the staggered module $\AffStagMod{0}$.  We indicate the composition factors $\tfunc{\gamma^{-2}}{\AffIrrMod{1}}$, $\AffIrrMod{0}$ and $\tfunc{\gamma^2}{\AffIrrMod{1}}$ with dashed boundary lines (left) and the multiplicities of a few of the weight spaces (right).  The latter are separated into those for each composition factor (the order is as above) with the multiplicities for $\AffIrrMod{0}$ doubled (as it appears twice as a factor).  At left, we split the weight spaces to distinguish the two $\AffIrrMod{0}$ factors (as in \figref{figEEGrade101}, the $L_0$-eigenstates appear above their Jordan partners).  We also indicate with arrows a few fundamental actions which define the indecomposable structure.} \label{figS0}
\end{center}
\end{figure}
}

It is well known that staggered modules for the Virasoro algebra need not be completely determined by their exact sequences \cite{GabInd96,RohRed96,RidSta09}.  Indeed, one must in general compute (up to two) additional numerical invariants, called beta-invariants or logarithmic couplings, which completely specify the module given an exact sequence \cite{RidPer07,RidSta09}.  We should therefore analyse the corresponding situation for our staggered $\AKMA{sl}{2}$-module $\AffStagMod{0}$.  Referring to \figref{figS0}, we see that the $\AKMA{sl}{2}$ action on $\ket{y_0}$ (and hence the $\AKMA{sl}{2}$ action on $\AffStagMod{0}$) will be determined once we have specified $e_0 \ket{y_0}$, $e_1 \ket{y_0}$, $f_0 \ket{y_0}$ and $f_1 \ket{y_0}$.

Let us recall our state definitions:
\begin{itemize}
\item First, choose $\ket{x_0^+}$ to be a state of $\SLA{sl}{2}$-weight $2$ and conformal dimension $-1$.  Declaring it to have norm $1$ defines the scalar product on the submodule $\AKMA{sl}{2} \ket{x_0^+} \cong \tfunc{\gamma^{-1}}{\AffOthMod{-1/2}^+}$ generated by $\ket{x_0^+}$.
\item Define $\ket{\omega_0} = f_{-1} \ket{x_0^+}$ and note that $e_1 \ket{\omega_0} = 0$ implies that $\braket{\omega_0}{\omega_0} = \bracket{\omega_0}{f_{-1}}{x_0^+} = 0$.  The $\AffIrrMod{0}$-submodule of $\AffStagMod{0}$ generated by $\ket{\omega_0}$ is therefore null.
\item Define $\ket{x_0^-}$ by imposing $e_{-1} \ket{x_0^-} = \ket{\omega_0}$.  Since $\ket{x_0^-} \notin \AKMA{sl}{2} \ket{x_0^+}$, we may suppose that $\ket{x_0^-}$ is also normalised.  This then defines the scalar product on the submodule $\AKMA{sl}{2} \ket{x_0^-} \cong \tfunc{\gamma^{-1}}{\AffOthMod{1/2}^-}$.  Note that both of these scalar products agree (indeed, they both vanish) on the intersection of the submodules $\AKMA{sl}{2} \ket{x_0^+} \cap \AKMA{sl}{2} \ket{x_0^-} = \AKMA{sl}{2} \ket{\omega_0} \cong \AffIrrMod{0}$.
\item Finally, let $\ket{y_0}$ be a state of $\SLA{sl}{2}$-weight $0$ and (generalised) conformal dimension $0$ satisfying $L_0 \ket{y_0} = \ket{\omega_0}$.  This only defines $\ket{y_0}$ up to adding arbitrary multiples of $\ket{\omega_0}$.  We cannot normalise $\ket{y_0}$.
\end{itemize}
From the multiplicities of \figref{figS0}, we can write
\begin{equation} \label{eqnDefBeta0}
e_1 \ket{y_0} = \beta_0 \ket{x^+_0} \qquad \text{and} \qquad e_0 \ket{y_0} = \brac{\beta_0' h_{-1} + \beta_0'' f_{-2} e_1} \ket{x^+_0},
\end{equation}
where $\beta_0$, $\beta_0'$ and $\beta_0''$ are unknown constants.  Note that a redefinition of $\ket{y_0}$ through adding some multiple of $\ket{\omega_0}$ would not affect the values of these constants.  There are, in principle, three similar constants defining the action of $f_1$ and $f_0$:
\begin{equation}
f_1 \ket{y_0} = \tilde{\beta}_0 \ket{x^-_0} \qquad \text{and} \qquad f_0 \ket{y_0} = \brac{\tilde{\beta}_0' h_{-1} + \tilde{\beta}_0'' e_{-2} f_1} \ket{x^-_0}.
\end{equation}
However, $\tilde{\beta}_0 = \bracket{x_0^-}{f_1}{y_0} = \braket{\omega_0}{y_0} = \bracket{x_0^+}{e_1}{y_0} = \beta_0$ and one can similarly use the scalar product to show that
\begin{equation}
2 \beta_0 - \tilde{\beta}_0' + 3 \tilde{\beta}_0'' = 0 \qquad \text{and} \qquad -2 \tilde{\beta}_0' + 3 \tilde{\beta}_0'' = 0.
\end{equation}
The $\AKMA{sl}{2}$-action is therefore defined by the three numbers $\beta_0$, $\beta_0'$ and $\beta_0''$.

However, these three unknown constants are themselves not independent, because
\begin{subequations} \label{eqnBeta0Cons}
\begin{align}
h_1 e_0 \ket{y_0} &= 2 e_1 \ket{y_0} & &\Rightarrow & 2 \beta_0 + \beta_0' + 3 \beta_0'' &= 0 \\
\text{and} \qquad e_2 e_0 \ket{y_0} &= 0 & &\Rightarrow & 2 \beta_0' + 3 \beta_0'' &= 0.
\end{align}
It now follows that $\tilde{\beta}_0' = -\beta_0'$ and $\tilde{\beta}_0'' = \beta_0''$, exactly as one would expect from applying the conjugation automorphism $\mathsf{w}$ to \eqnref{eqnDefBeta0}.  Moreover, one may check that $e_1 e_0 \ket{y_0} = e_0 e_1 \ket{y_0}$ leads to two constraints which are not independent of those given in \eqref{eqnBeta0Cons}.  Note that these constraints are all homogeneous.  In contrast, the normalisation that we chose above for the Jordan partner gives an additional inhomogeneous constraint:
\begin{equation}
L_0 \ket{y_0} = \ket{\omega_0} \qquad \Rightarrow \qquad 4 \beta_0 + 4 \beta_0' = -3.
\end{equation}
\end{subequations}
We can therefore solve the three independent constraints \eqref{eqnBeta0Cons} to obtain
\begin{equation}
\beta_0 = -\frac{1}{4}, \qquad \beta_0' = -\frac{1}{2} \qquad \text{and} \qquad \beta_0'' = \frac{1}{3}.
\end{equation}
These numbers uniquely determine the $\AKMA{sl}{2}$-action on $\AffStagMod{0}$.

\subsection{Analysis when $\lambda + \mu = 1$} \label{secAnalysis2}

Combining the associativity of the fusion rules with \eqnDref{eqnLEFusion2}{eqnEEFus0}, we obtain 
\begin{equation} \label{eqnEEFus1}
\AffOthMod{\lambda} \fuse \AffOthMod{\mu} = \AffIrrMod{1} \fuse \AffStagMod{0} \equiv \AffStagMod{1} \qquad \text{if $\lambda + \mu = 1 \pmod{2}$,}
\end{equation}
which defines the $\AKMA{sl}{2}$-module $\AffStagMod{1}$.  As we have exhibited $\AffStagMod{0}$ as an indecomposable combination of $\tfunc{\gamma^2}{\AffIrrMod{1}}$, $\tfunc{\gamma^{-2}}{\AffIrrMod{1}}$ and two copies of $\AffIrrMod{0}$, it is very natural to presume that $\AffStagMod{1}$ may be likewise exhibited as an indecomposable combination of $\tfunc{\gamma^2}{\AffIrrMod{0}}$, $\tfunc{\gamma^{-2}}{\AffIrrMod{0}}$ and two copies of $\AffIrrMod{1}$.  Indeed, comparing \figref{figSpecFlow} with the fusion results pictured in \figref{figEEGrade101} (right), we see that this presumption is supported except in that we only see one copy of $\AffIrrMod{1}$.

The explanation for the missing copy of $\AffIrrMod{1}$ is much the same as for the missing descendants of $\ket{\omega_0}$ in the previous section.  First, our presumption above for the structure of $\AffStagMod{1}$ lets us choose a state $\ket{x^+_1}$ of $\SLA{sl}{2}$-weight $1$ and conformal dimension $-\tfrac{1}{2}$.  This generates the copy of $\tfunc{\gamma^{-2}}{\AffIrrMod{0}}$, or more accurately, it generates an indecomposable module $\tfunc{\gamma^{-1}}{\AffOthMod{-3/2}^+}$ defined by the exact sequence
\begin{equation}
\dses{\AffIrrMod{1}}{\tfunc{\gamma^{-1}}{\AffOthMod{-3/2}^+}}{\tfunc{\gamma^{-2}}{\AffIrrMod{0}}}.
\end{equation}
The \hws{} of the $\AffIrrMod{1}$-submodule is then
\begin{equation}
\ket{\omega^+_1} = -e_0 f_{-1} \ket{x^+_1}.
\end{equation}
Substituting $m = \tfrac{1}{2}$ into the vanishing $\AffOthMod{}$-type singular vector \eqref{eqnSV1E} and applying $\gamma^{-1}$ now gives
\begin{equation} \label{eqnOmega1}
\ket{\omega^+_1} = \frac{1}{2} \brac{h_{-1} - f_{-2} e_1} \ket{x^+_1},
\end{equation}
which shows that this copy of $\AffIrrMod{1}$ would not be uncovered in a fusion computation to grade $\left(1,0,1\right)$ in accord with what we have observed in \secref{secFusEEAgain}.

Of course, this is currently pure supposition, if rather well-founded.  We have not yet managed to observe a non-diagonalisable action of $L_0$ on $\AffStagMod{1}$.  However, \eqnref{eqnOmega1} indicates how this can be achieved:  We can simply exclude $h_{-1}$ from the algebra $\alg{A}$ controlling the fusion algorithm.  With such an algebra, both $\ket{\omega^+_1}$ and its Jordan partner $\ket{y^+_1}$ should be visible.  We therefore expect that this slight change to $\alg{A}$ will enable us to detect a Jordan structure for $L_0$.

We have therefore repeated the fusion computation for $\AffOthMod{\lambda} \fuse \AffOthMod{\mu}$ one last time, taking the algebra $\alg{A}$ to be generated by the $e_{-n}$, $h_{-n}$ and $f_{-n}$ with $n \geqslant 2$ \emph{and} the $h_{-1}^j$ with $j \geqslant 2$.  This may seem like a small change, but the corresponding increase in algorithmic complexity is significant.  The part of the fusion module that this uncovers may be found within the space spanned by the states
\begin{equation}
\ket{v_n} \otimes \ket{v_m}, \qquad \ket{v_n} \otimes h_{-1} \ket{v_m}, \qquad \ket{v_n} \otimes e_{-1} \ket{v_{m-2}} \qquad \text{and} \qquad \ket{v_n} \otimes h_{-1} e_{-1} \ket{v_{m-2}}.
\end{equation}
Using the vanishing vectors \eqref{eqnSV1E} and \eqref{eqnSV2E} of $\AffOthMod{\lambda}$, we derive four recursion relations which together bound the dimension of the $\SLA{sl}{2}$-weight spaces by $12$.  Further analysis reduces this to $6$ when the weight is $\pm 1$ and $4$ in general.  The explicit construction of $L_0$ confirms that it indeed has a non-trivial Jordan cell corresponding to eigenvalue $\tfrac{1}{2}$ when the $\SLA{sl}{2}$-weight is $\pm 1$.  For completeness, we illustrate the results of this fusion computation in \figref{figS1-fus}.  It is not hard to check that these results completely support the structure of $\AffStagMod{1}$ proposed above.

{
\psfrag{y+}[][]{$\ket{y_1^+}$}
\psfrag{y-}[][]{$\ket{y_1^-}$}
\psfrag{w+}[][]{$\ket{\omega_1^+}$}
\psfrag{w-}[][]{$\ket{\omega_1^-}$}
\psfrag{x+}[][]{$\ket{x_1^+}$}
\psfrag{x-}[][]{$\ket{x_1^-}$}
\begin{figure}
\begin{center}
\includegraphics[width=10cm]{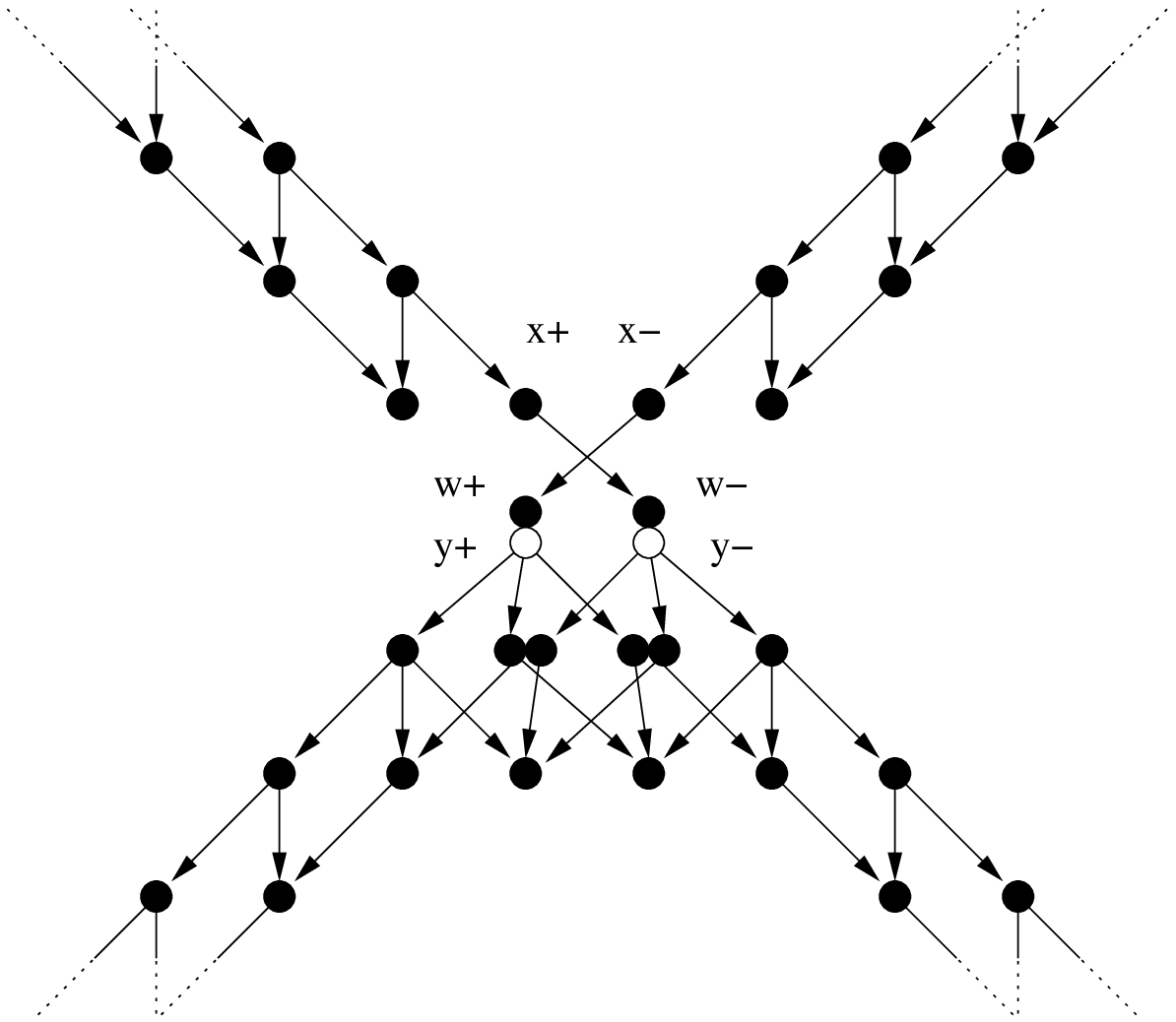}
\caption{A schematic illustration of the states of the quotient of the fusion module $\AffOthMod{0} \fuse \AffOthMod{1}$ when the algebra $\alg{A}$ is generated by the $e_{-n}$, $h_{-n}$ and $f_{-n}$ with $n \geqslant 2$ and the $h_{-1}^j$ with $j \geqslant 2$.  Again, we only indicate the actions of $e_{-1}$, $h_{-1}$ and $f_{-1}$.  Weight spaces carrying a non-diagonalisable action are indicated by arranging the states vertically (the white states are the Jordan partners).  Also noted are the states $\ket{y_1^{\pm}}$, $\ket{\omega_1^{\pm}}$ and $\ket{x_1^{\pm}}$.} \label{figS1-fus}
\end{center}
\end{figure}
}

It is now appropriate to ask if the $\AKMA{sl}{2}$-action on $\AffStagMod{1}$ is uniquely determined by its structure, or if there are additional logarithmic couplings to compute.  We define states in $\AffStagMod{1}$ as follows:
\begin{itemize}
\item Choose $\ket{x_1^+}$ to be a state of $\SLA{sl}{2}$-weight $1$ and conformal dimension $-\tfrac{1}{2}$.  We define the scalar product on the submodule $\AKMA{sl}{2} \ket{x_1^+} \cong \tfunc{\gamma^{-1}}{\AffOthMod{-3/2}^+}$ by declaring that $\ket{x_1^+}$ has norm $1$.
\item Define $\ket{\omega_1^-} = f_{-1} \ket{x_1^+}$ and $\ket{\omega_1^+} = -e_0 \ket{\omega_1^-}$.  Then, $f_0 \ket{\omega_1^+} = \ket{\omega_1^-}$ and the $\AffIrrMod{1}$-submodule generated by $\ket{\omega_1^+}$ consists entirely of zero-norm states.
\item Define $\ket{x_1^-}$ by setting $e_{-1} \ket{x_1^-} = \ket{\omega_1^+}$.  The scalar product on $\AKMA{sl}{2} \ket{x_1^-} \cong \tfunc{\gamma}{\AffOthMod{3/2}^-}$ is then determined by defining the norm of $\ket{x_1^-}$ to be $1$.  Again, these scalar products agree (they both vanish) on the intersection $\AKMA{sl}{2} \ket{x_1^+} \cap \AKMA{sl}{2} \ket{x_1^-} \cong \AffIrrMod{1}$.
\item Choose $\ket{y_1^+}$ to be a state of $\SLA{sl}{2}$-weight $1$ and conformal dimension $\tfrac{1}{2}$ that satisfies $\brac{L_0 - \tfrac{1}{2}} \ket{y_1^+} = \ket{\omega_1^+}$.  Then, define $\ket{y_1^-} = f_0 \ket{y_1^+}$ so that $\brac{L_0 - \tfrac{1}{2}} \ket{y_1^-} = \ket{\omega_1^-}$.  The $\ket{y_1^{\pm}}$ are not normalisable.  Note that $\ket{y_1^+}$ is only defined up to adding arbitrary multiples of $\ket{\omega_1^+}$ \emph{and} $h_{-1} \ket{x_1^+}$.
\end{itemize}
We illustrate the structure of $\AffStagMod{1}$ and the multiplicity of some of its weight spaces in \figref{figS1}.  The latter multiplicities make it clear that there are \emph{ten} unknown constants which define the action of $\AKMA{sl}{2}$ on $\AffStagMod{1}$.\footnote{Actually, it is clear from the outset that knowing the action of $e_0$ and $f_1$ is sufficient in this case.  We shall, however, ignore this slight simplification for pedagogical reasons.}  We let
\begin{subequations} \label{eqnS1DefBetas}
\begin{gather}
h_1 \ket{y_1^+} = \beta \ket{x_1^+}, \quad f_1 \ket{y_1^+} = \beta_1 \ket{x_1^-}, \quad e_2 \ket{y_1^+} = \beta_2 e_1 \ket{x_1^+}, \quad e_1 \ket{y_1^+} = \brac{\gamma_1 e_0 + \gamma_2 h_{-1} e_1} \ket{x_1^+}, \\
\text{and} \qquad e_0 \ket{y_1^+} = \brac{\alpha_1 e_{-1} + \alpha_2 h_{-1} e_0 + \alpha_3 h_{-2} e_1 + \alpha_4 h_{-1}^2 e_1 + \alpha_5 f_{-3} e_1^2} \ket{x_1^+}.
\end{gather}
\end{subequations}
As in \secref{secAnalysis1}, these constants are not all independent.

{
\psfrag{L0}[][]{$\AffIrrMod{1}$}
\psfrag{S0}[][]{$\AffStagMod{1}$}
\psfrag{E+}[][]{$\tfunc{\gamma^{-2}}{\AffIrrMod{0}}$}
\psfrag{E-}[][]{$\tfunc{\gamma^2}{\AffIrrMod{0}}$}
\setlength{\extrarowheight}{6mm}
\begin{figure}
\begin{center}
\parbox{7.5cm}{
\includegraphics[width=7cm]{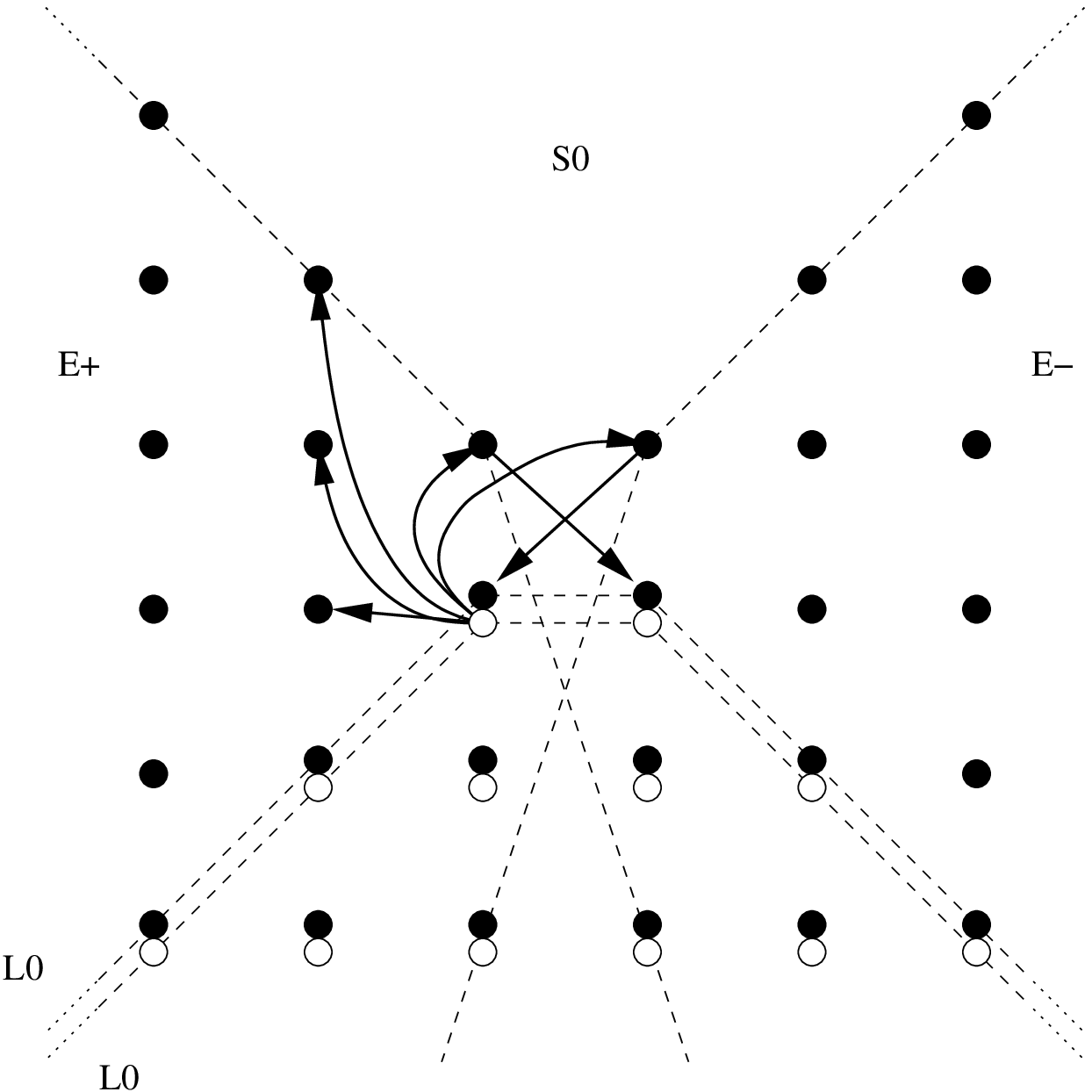}
}
\hspace{10mm}
\parbox{5.5cm}{
\begin{tabular}{*{4}{@{}C@{\hspace{2mm}}}}
\brac{1,0,0} & & & \brac{0,0,1} \\
\brac{2,0,0} & \brac{1,0,0} & \brac{0,0,1} & \brac{0,0,2} \\
\brac{5,0,0} & \brac{1,2,0} & \brac{0,2,1} & \brac{0,0,5} \\
\brac{9,2,0} & \brac{3,4,0} & \brac{0,4,3} & \brac{0,2,9} \\
\brac{18,4,0} & \brac{6,8,1} & \brac{1,8,6} & \brac{0,4,18}
\end{tabular}
}
\caption{Schematic illustrations of the states and multiplicities of the staggered module $\AffStagMod{1}$, following the conventions established for \figref{figS0}.  The multiplicities are for the composition modules $\tfunc{\gamma^{-2}}{\AffIrrMod{0}}$, $\AffIrrMod{1}$ (doubled) and $\tfunc{\gamma^2}{\AffIrrMod{0}}$.  As before, we indicate with arrows a few fundamental actions which result from the indecomposable structure.} \label{figS1}
\end{center}
\end{figure}
}

We proceed by considering the effect of combining these definitions with the commutation relations.  For example, we can evaluate $e_1 e_2 \ket{y_1^+}$ in two ways, leading to
\begin{subequations} \label{eqnS1Cons}
\begin{equation}
\beta_2 e_1^2 \ket{x_1^+} = e_1 e_2 \ket{y_1^+} = e_2 e_1 \ket{y_1^+} = \gamma_1 e_2 e_0 \ket{x_1^+} + \gamma_2 e_2 h_{-1} e_1 \ket{x_1^+} = -2 \gamma_2 e_1^2 \ket{x_1^+},
\end{equation}
hence $\beta_2 + 2 \gamma_2 = 0$.  Similarly, considering the action of $e_1 h_1$, $e_0 h_1$, $e_0 f_1$, $h_2 e_0$ and $e_3 e_0$ on $\ket{y_1^+}$ leads to six more independent homogeneous constraints:
\begin{align}
\beta + 2 \beta_2 - 2 \gamma_1 + \gamma_2 &= 0, & \gamma_2 - \alpha_2 + \alpha_4 + 2 \alpha_5 &= 0, & \beta + 2 \gamma_1 - 2 \alpha_1 + \alpha_2 - 4 \alpha_5 &= 0, \\
\beta_2 - \alpha_1 + \alpha_3 + 3 \alpha_5 &= 0, & 4 \alpha_3 - 8 \alpha_4 + 7 \alpha_5 &= 0, & 2 \beta - 3 \alpha_1 - 4 \alpha_2 - 2 \alpha_3 - 4 \alpha_4 &= 0.
\end{align}
Again, the Jordan structure leads to inhomogeneous constraints.  Expanding $L_0 \ket{y_1^+} = \tfrac{1}{2} \ket{y_1^+} + \ket{\omega_1^+}$, for example, leads to
\begin{equation}
-2 \beta_1 + 4 \beta_2 + 2 \gamma_1 + 8 \gamma_2 + 4 \alpha_2 + 8 \alpha_3 + 16 \alpha_4 = 3.
\end{equation}
Repeating this for $\ket{y_1^-}$ and $\ket{\omega_1^-}$ leads to another independent inhomogeneous constraint:
\begin{equation} \label{eqnS1Last}
2 \beta + 6 \beta_1 - 2 \gamma_1 - 2 \gamma_2 - 2 \alpha_1 - 4 \alpha_2 - 4 \alpha_3 = 3.
\end{equation}
\end{subequations}
However, its derivation requires a significant digression.  As $\ket{y_1^-}$ has been defined to be $f_0 \ket{y_1^+}$, it is straight-forward to obtain
\begin{subequations} \label{eqnS1DefBetas'}
\begin{equation}
h_1 \ket{y_1^-} = -2 \beta_1 \ket{x_1^-}, \quad e_1 \ket{y_1^-} = \brac{\gamma_1 + \gamma_2 - \beta} \ket{x_1^+}, \quad f_2 \ket{y_1^-} = 0 \quad \text{and} \quad f_1 \ket{y_1^-} = \beta_1 f_0 \ket{x_1^-}.
\end{equation}
Determining $f_0 \ket{y_1^-}$, however, requires more of the games that led to the constraints \eqref{eqnS1Cons}.  We omit the details and give only the result:
\begin{equation}
f_0 \ket{y_1^-} = \frac{4 \beta_1}{15} \brac{3 f_{-1} - 7 h_{-1} f_0 + 9 h_{-2} f_1 - h_{-1}^2 f_1 + 4 e_{-3} f_1^2} \ket{x_1^-}.
\end{equation}
\end{subequations}
With this, \eqref{eqnS1Last} is easily derived.

We therefore have nine constraint equations in ten unknowns and these are all the constraints that one can find.  This does not mean that there is a one-parameter family of modules that form candidates for $\AffStagMod{1}$.  Rather, it reflects the fact that we can only choose $\ket{y_1^+}$ up to multiples of $\ket{\omega_1^+}$ and $h_{-1} \ket{x_1^+}$.  As $\ket{\omega_1^+}$ is annihilated by $h_1$, $f_1$, $e_2$, $e_1$ and $e_0$, a redefinition of the form
\begin{equation}
\ket{y_1^+} \longmapsto \ket{y_1^+} + \alpha \ket{\omega_1^+}
\end{equation}
does not change the constants $\beta$, $\beta_i$, $\gamma_i$ and $\alpha_i$ which we defined in \eqnref{eqnS1DefBetas}.  A redefinition of the form
\begin{equation}
\ket{y_1^+} \longmapsto \ket{y_1^+} + \alpha h_{-1} \ket{x_1^+},
\end{equation}
however, will change some of these constants, specifically $\beta$, $\beta_2$, $\gamma_1$, $\gamma_2$, $\alpha_1$ and $\alpha_2$.  This is reflected in the general solution to the constraints \eqref{eqnS1Cons}:
\begin{subequations}
\begin{align}
\beta &= -\frac{29}{30} - \alpha, & \beta_1 &= \frac{1}{4}, & \beta_2 &= -\frac{14}{15} - 2 \alpha, & \gamma_1 &= -\frac{71}{60} - 2 \alpha, & \gamma_2 &= \frac{7}{15} + \alpha, \\
\alpha_1 &= -\frac{17}{15} - 2 \alpha, & \alpha_2 &= \alpha, & \alpha_3 &= \frac{3}{5}, & \alpha_4 &= \frac{1}{15}, & \alpha_5 &= -\frac{4}{15}.
\end{align}
\end{subequations}
In the language of \cite{RidPer07}, the constants $\beta$, $\beta_2$, $\gamma_1$, $\gamma_2$, $\alpha_1$ and $\alpha_2$ are not \emph{gauge-invariant}.

We conclude this analysis by remarking that a quick comparison of \eqnDref{eqnS1DefBetas}{eqnS1DefBetas'} shows that, unlike that of \secref{secAnalysis1}, our analysis has not been invariant under the conjugation automorphism $\mathsf{w}$.  The reason can be traced back to the definition $\ket{y_1^-} = f_0 \ket{y_1^+}$.  This breaks conjugation-invariance rather badly because
\begin{equation}
e_0 \ket{y_1^-} = -\ket{y_1^+} + f_0 e_0 \ket{y_1^+} = -\ket{y_1^+} + \Bigl( \frac{7}{15} + \alpha \Bigr) h_{-1} \ket{x_1^+} - \Bigl( \frac{44}{15} + 2 \alpha \Bigr) \ket{\omega_1^+},
\end{equation}
rather than just $e_0 \ket{y_1^-} = -\ket{y_1^+}$.  It would be nice to correct this, but we feel that the complexity that this would add is rather unjustified at present.

\subsection{Fusing $\AffOthMod{\lambda}$ and $\AffStagMod{\mu}$} \label{secFusES}

It remains to compute the fusion rules involving these new staggered modules $\AffStagMod{0}$ and $\AffStagMod{1}$.  Associativity and \eqnref{eqnEEFus1} show that
\begin{equation}
\AffIrrMod{\lambda} \fuse \AffStagMod{\mu} = \AffStagMod{\lambda + \mu},
\end{equation}
so our next task is to determine the fusion of $\AffOthMod{\lambda}$ and $\AffStagMod{\mu}$.  Performing the fusion algorithm with staggered modules is not an easy task, especially in view of the rather involved structure of $\AffStagMod{1}$.  Luckily, associativity again reduces the burden somewhat --- the fusions with $\AffStagMod{1}$ will follow once we know those with $\AffStagMod{0}$.  We therefore turn to the computation of $\AffOthMod{\lambda} \fuse \AffStagMod{0}$ to grade $0$.

We first need to decide on an appropriate tensor product space in which to find the grade $0$ fusion product.  The general theory suggests that we should consider the space spanned by the zero-grade states of $\AffOthMod{\lambda}$ tensored with the zero-grade states of $\AffStagMod{0}$.  The latter are those states which are not in $\alg{A}^- \AffStagMod{0}$ (recall that we defined the algebra $\alg{A}^-$ in \secref{secFus0Pre}), hence cannot be written as $J_{-n} \ket{w}$ for some $J=e,h,f$, $n>0$ and $\ket{w} \in \AffStagMod{0}$.  A little reflection shows that only $\ket{y_0}$ has this property, hence we should consider the space spanned by the
\begin{equation} \label{eqnFirstTry}
\ket{v_m} \otimes \ket{y_0} \qquad \text{($m \in 2 \ZZ + \lambda$).}
\end{equation}
However, it seems that this cannot be the right space as it is not clear how to use \eqnDref{eqnStep1}{eqnStep2} to reduce $\ket{v_{m-2}} \otimes \ket{x_0^+}$, say, to something proportional to $\ket{v_m} \otimes \ket{y_0}$.

Instead, we can try the space spanned by
\begin{equation} \label{eqnTry}
\ket{v_{m-2n}} \otimes e_1^{n-1} \ket{x_0^+}, \qquad \ket{v_m} \otimes \ket{y_0} \qquad \text{and} \qquad \ket{v_{m+2n}} \otimes f_1^{n-1} \ket{x_0^-} \qquad \text{($m \in 2 \ZZ + \lambda$, $n \in \ZZ_+$).}
\end{equation}
Applying spectral flow to the vanishing singular vectors \eqref{eqnSV1E}, we obtain a relation expressing $e_0 e_1^{n-1} \ket{x_0^+}$ as a linear combination of $h_{-1} e_1^n \ket{x_0^+}$ and $f_{-2} e_1^{n+1} \ket{x_0^+}$.  There is a similar relation for $f_0 f_1^{n-1} \ket{x_0^-}$, hence we see that the states complementary to the span of \eqref{eqnTry} can all be expressed as (linear combinations of the) $J_{-n} \ket{w}$ with $J=e,h,f$ and $n>0$.  It is now easy to see that the procedure of \secref{secFus0Pre} terminates, hence that the span of \eqref{eqnTry} contains the grade $0$ fusion product.

Note however that $\func{\Delta}{e_{-1}} = 0$ implies that
\begin{equation} \label{eqnCons1}
\frac{n \brac{2n+1}}{2} \ket{v_m} \otimes f_1^{n-1} \ket{x_0^-} = \ket{v_m} \otimes e_{-1} f_1^n \ket{x_0^-} = -e_0 \ket{v_m} \otimes f_1^n \ket{x_0^-} = -\ket{v_{m+2}} \otimes f_1^n \ket{x_0^-},
\end{equation}
hence that every $\ket{v_m} \otimes f_1^n \ket{x_0^-}$ is proportional to $\ket{v_{m-2n}} \otimes \ket{x_0^-}$.  Similarly,
\begin{equation} \label{eqnCons2}
\frac{n \brac{2n+1}}{2} \ket{v_m} \otimes e_1^{n-1} \ket{x_0^+} = -\frac{\brac{2m-1} \brac{2m-3}}{16} \ket{v_{m-2}} \otimes e_1^n \ket{x_0^+},
\end{equation}
so $\ket{v_m} \otimes e_1^n \ket{x_0^+}$ is seen to be proportional to $\ket{v_{m+2n}} \otimes \ket{x_0^+}$.  Moreover, the same manipulations give
\begin{equation} \label{eqnCons3}
-\ket{v_{m+2}} \otimes \ket{x_0^-} = \ket{v_m} \otimes \ket{\omega_0} = \frac{\brac{2m-1} \brac{2m-3}}{16} \ket{v_{m-2}} \otimes \ket{x_0^+}.
\end{equation}
The upshot is that the rather large space spanned by the vectors of \eqref{eqnTry} may be replaced by the span of
\begin{equation} \label{eqnGotIt}
\ket{v_m} \otimes \ket{y_0} \qquad \text{and} \qquad \ket{v_m} \otimes \ket{\omega_0}.
\end{equation}

We have not yet used the vanishing singular vectors to search for spurious states.  We have therefore coupled the vanishing vectors \eqref{eqnSV1E} and \eqref{eqnSV2E} of $\AffOthMod{\lambda}$ to $\ket{\omega_0}$ and $\ket{y_0}$, but find nothing.  We have also checked that the vanishing vector \eqref{eqnSVL0} of $\AffIrrMod{0} \subset \AffStagMod{0}$ yields no spurious states.\footnote{There is also the vanishing singular vector obtained from \eqref{eqnSVL0} by replacing $\ket{0}$ by $\ket{y_0}$ and adding certain terms from $\AKMA{sl}{2} \ket{x_0^+} + \AKMA{sl}{2} \ket{x_0^-}$.  We did not check this singular vector as determining these extra terms did not seem worth the trouble.}  The space spanned by the vectors of \eqref{eqnGotIt} therefore seems to give the correct grade $0$ fusion product.  The $\SLA{sl}{2}$-action may be checked to be that of the zero-grade subspace of two copies of $\AffOthMod{\lambda}$ and we compute that
\begin{equation}
\func{\Delta}{L_0} = 
\begin{pmatrix}
-\tfrac{1}{8} & 0 \\
0 & -\tfrac{1}{8}
\end{pmatrix}
\end{equation}
with respect to the ordered basis \eqref{eqnGotIt}.  This suggests that
\begin{equation} \label{eqnFusES0'}
\AffOthMod{\lambda} \fuse \AffStagMod{0} = 2 \AffOthMod{\lambda}.
\end{equation}

To confirm this, we have repeated the fusion computation to grade $\brac{0,0,1}$, meaning that we take the algebra $\alg{A}$ of \secref{secFusPre} to be that generated by the $e_{-n}$, $h_{-n}$ and $f_{-n-1}$ with $n \geqslant 1$.  This time, we consider the span of the
\begin{equation}
\ket{v_{m-2n}} \otimes e_1^{n-1} \ket{x_0^+}, \qquad \ket{v_{m+2n}} \otimes f_{-1}^n \ket{y_0} \qquad \text{and} \qquad \ket{v_{m+2n}} \otimes f_1^{n-1} \ket{x_0^-} \qquad \text{($m \in 2 \ZZ + \lambda$).}
\end{equation}
Applying spectral flow to the vanishing singular vectors \eqref{eqnSV1E} and \eqref{eqnSV2E} allows us to deal with $e_0 e_1^{n-1} \ket{x_0^+}$, $f_0 f_1^{n-1} \ket{x_0^-}$ and $f_{-1} f_1^{n-1} \ket{x_0^-}$ as before.  Again, \eqnref{eqnCons1} and the first equality of \eqnref{eqnCons3} allow us to replace this space by the span of the
\begin{equation}
\ket{v_{m-2n}} \otimes e_1^{n-1} \ket{x_0^+}, \qquad \ket{v_{m+2n}} \otimes f_{-1}^n \ket{y_0} \qquad \text{and} \qquad \ket{v_m} \otimes \ket{\omega_0} \qquad \text{($m \in 2 \ZZ + \lambda$).}
\end{equation}
A little computation now shows that the singular vectors of $\AffOthMod{\lambda}$ then reduce the grade $\brac{0,0,1}$ fusion to the span of the vectors of \eqref{eqnGotIt}, confirming \eqnref{eqnFusES0'}.  More precisely, this (and the fact that the fusion product must be self-conjugate) rules out the twisted modules $\tfunc{\gamma^{\pm 2}}{\AffIrrMod{\mu}}$ and $\tfunc{\gamma^{\pm 1}}{\AffOthMod{\mu}}$ as composition factors of the result.

However, we have reason to suspect that \eqnref{eqnFusES0'} is incorrect, though we shall not elaborate on why until the next section.  Suffice to say that we have also computed the fusion to grade $\brac{0,0,2}$, so that $\alg{A}$ is generated by the $e_{-n}$, $h_{-n}$ and $f_{-n-2}$ with $n \geqslant 1$.  A careful analysis along the lines of the previous analyses shows that the fusion product lies within the span of the vectors
\begin{subequations}
\begin{gather}
\ket{v_{m-2n}} \otimes e_1^{n-1} \ket{x_0^+}, \qquad \ket{v_{m-2n}} \otimes f_{-2} e_1^n \ket{x_0^+}, \\
\ket{v_{m+2n}} \otimes f_{-1}^n \ket{y_0}, \qquad \ket{v_{m+2n}} \otimes f_{-2} f_{-1}^{n-1} \ket{y_0}, \qquad \text{and} \qquad \ket{v_m} \otimes \ket{\omega_0}.
\end{gather}
\end{subequations}
This time, the singular vectors of $\AffOthMod{\lambda}$ reduce the grade $\brac{0,0,2}$ fusion to the span of the vectors
\begin{equation}
\ket{v_{m-2}} \otimes \ket{x_0^+}, \qquad \ket{v_m} \otimes \ket{y_0} \qquad \text{and} \qquad \ket{v_m} \otimes \ket{\omega_0} \qquad \text{($m \in 2 \ZZ + \lambda$).}
\end{equation}
Explicitly computing the eigenvalues of $\func{\Delta}{L_0}$ gives $-\tfrac{1}{8}$, $-\tfrac{1}{8}$ and $-m + \tfrac{3}{8}$, confirming our suspicion that \eqnref{eqnFusES0} is not quite right.  Rather, coupling this result with the requirement that the result be invariant under conjugation leads to
\begin{equation} \label{eqnFusES0}
\AffOthMod{\lambda} \fuse \AffStagMod{0} = \tfunc{\gamma^{-2}}{\AffOthMod{\lambda + 1}} \oplus 2 \AffOthMod{\lambda} \oplus \tfunc{\gamma^2}{\AffOthMod{\lambda + 1}}.
\end{equation}
This rule is of course conjectural, though we will discuss in the next section why we are confident that it is indeed correct.  The fact that there is no indecomposable structure involving $\AffOthMod{\lambda}$ and the twisted modules follows from the difference between the fractional parts of the conformal dimensions of the states of these irreducibles.  However, we have not ruled out the presence of further twisted modules in the above decomposition, nor the possibility that the $\AffOthMod{\lambda}$ or the twisted $\AffOthMod{\lambda + 1}$ are composition factors of an indecomposable.  We view this as unlikely, but settling this completely would require further computations along the lines of those presented above, or some abstract mathematical results generalising those of \cite{RidSta09} to $\AKMA{sl}{2}$-modules.

\section{Summary of Results and Discussion} \label{secSummary}

The results derived in \secDref{secFusion0}{secFusion} give, when coupled with associativity, the fusion rings of the $\AKMA{sl}{2}_{-1/2}$ theories considered in \secref{secOld}.  The spectrum consists of four irreducible untwisted $\AKMA{sl}{2}$-modules $\AffIrrMod{0}$, $\AffIrrMod{1}$, $\AffOthMod{0}$ and $\AffOthMod{1}$, two indecomposable untwisted modules $\AffStagMod{0}$ and $\AffStagMod{1}$, and their twisted versions under the spectral flow automorphism $\gamma$.  The fusion rules themselves can be put in a compact form by using their (conjectured) covariant behaviour under $\gamma$ (\eqnref{eqnFusionAssumption}).  This allows us to restrict to the untwisted sector in which the fusion rules are
\begin{align} \label{eqnFusion}
&
\begin{aligned}
\AffIrrMod{\lambda} \fuse \AffIrrMod{\mu} &= \AffIrrMod{\lambda + \mu}, \\ \AffIrrMod{\lambda} \fuse \AffOthMod{\mu} &= \AffOthMod{\lambda + \mu}, \\ \AffIrrMod{\lambda} \fuse \AffStagMod{\mu} &= \AffStagMod{\lambda + \mu},
\end{aligned}
& &
\begin{aligned}
\AffOthMod{\lambda} \fuse \AffOthMod{\mu} &= \AffStagMod{\lambda + \mu}, \\ \AffOthMod{\lambda} \fuse \AffStagMod{\mu} &= \tfunc{\gamma^{-2}}{\AffOthMod{\lambda + \mu + 1}} \oplus 2 \AffOthMod{\lambda + \mu} \oplus \tfunc{\gamma^2}{\AffOthMod{\lambda + \mu + 1}}, \\
\AffStagMod{\lambda} \fuse \AffStagMod{\mu} &= \tfunc{\gamma^{-2}}{\AffStagMod{\lambda + \mu + 1}} \oplus 2 \AffStagMod{\lambda + \mu} \oplus \tfunc{\gamma^2}{\AffStagMod{\lambda + \mu + 1}},
\end{aligned}
\end{align}
where, as throughout, the addition of the indices is understood to be \emph{modulo} $2$.  These rules confirm the claim made in \cite{LesLog04} that fusion generates no further indecomposables.  However, no fusion rules were given there, so our results go well beyond what was previously known.

The fusion rules \eqref{eqnFusion} report that which was deduced from the explicit computation of the fusion product to certain grades, as described in \secDref{secFusion0}{secFusion}.  As such, we cannot always rule out the possibility that the true fusion product involves highly twisted composition factors which our analysis has missed.  However, we have been able to \emph{prove} that such factors are absent in the fusion rules involving $\AffIrrMod{\lambda}$.  It would be very useful to refine the argument of these proofs to rule out highly twisted modules for more general fusions.

In fact, the above results describe somewhat more.  The uniform description of the $\AffOthMod{\lambda}$ for $\lambda \notin \ZZ + \tfrac{1}{2}$ means that the results described in \secDref{secFusion0}{secFusion} are not only valid for $\lambda$ and $\mu$ integral.\footnote{Recall that when $\lambda \in \ZZ + \tfrac{1}{2}$, the $\AffOthMod{}$-type modules are no longer irreducible, so one might have to exclude them from this remark, or modify it appropriately.}  In particular, we can deduce that the rules \eqref{eqnFusion} hold more generally, except for the replacement
\begin{equation} \label{eqnFusEEFull}
\AffOthMod{\lambda} \fuse \AffOthMod{\mu} = 
\begin{cases}
\AffStagMod{\lambda + \mu} & \text{if $\lambda + \mu \in \ZZ$,} \\
\tfunc{\gamma}{\AffOthMod{\lambda + \mu + 1/2}} \oplus \tfunc{\gamma^{-1}}{\AffOthMod{\lambda + \mu - 1/2}} & \text{otherwise.}
\end{cases}
\end{equation}
That the sum is direct in this fusion rule follows from the fact that the fractional parts of the conformal dimensions of the states in the two factors do not agree (equivalently, the eigenvalues of the ``central'' element $e^{2 \pi \ii L_0}$ are different).  This fusion rule should be relevant to more general models with $\AKMA{sl}{2}_{-1/2}$-symmetry, such as the various compactifications/orbifolds of the $\beta \gamma$ ghost theories.

We have also completely determined the structure of the indecomposable modules $\AffStagMod{0}$ and $\AffStagMod{1}$.  In brief, $\AffStagMod{\lambda}$ is composed of four irreducibles, its composition factors
\begin{equation}
\AffIrrMod{\lambda}, \quad \tfunc{\gamma^{-2}}{\AffIrrMod{\lambda + 1}}, \quad \tfunc{\gamma^2}{\AffIrrMod{\lambda + 1}} \quad \text{and} \quad \AffIrrMod{\lambda},
\end{equation}
which are ``glued'' together into an indecomposable as follows:
\begin{center}
\begin{tikzpicture}[thick,
	nom/.style={circle,draw=black!20,fill=black!20,inner sep=1pt}
	]
\node (top0) at (0,1.5) [] {$\AffIrrMod{0}$};
\node (left0) at (-1.5,0) [] {$\tfunc{\gamma^{-2}}{\AffIrrMod{1}}$};
\node (right0) at (1.5,0) [] {$\tfunc{\gamma^2}{\AffIrrMod{1}}$};
\node (bot0) at (0,-1.5) [] {$\AffIrrMod{0}$};
\node (top1) at (6,1.5) [] {$\AffIrrMod{1}$};
\node (left1) at (4.5,0) [] {$\tfunc{\gamma^{-2}}{\AffIrrMod{0}}$};
\node (right1) at (7.5,0) [] {$\tfunc{\gamma^2}{\AffIrrMod{0}}$};
\node (bot1) at (6,-1.5) [] {$\AffIrrMod{1}$};
\node at (0,0) [nom] {$\AffStagMod{0}$};
\node at (6,0) [nom] {$\AffStagMod{1}$};
\draw [->] (top0) -- (left0);
\draw [->] (top0) -- (right0);
\draw [->] (left0) -- (bot0);
\draw [->] (right0) -- (bot0);
\draw [->] (top1) -- (left1);
\draw [->] (top1) -- (right1);
\draw [->] (left1) -- (bot1);
\draw [->] (right1) -- (bot1);
\end{tikzpicture}
.
\end{center}
The arrows in these diagrams indicate the ``direction'' of the $\AKMA{sl}{2}$-action.  For example, the composition factors appearing in the bottom row describe the unique irreducible submodules (the \emph{socles}) of the $\AffStagMod{\lambda}$.  We also see that each $\AffStagMod{\lambda}$ covers the corresponding irreducible $\AffIrrMod{\lambda}$ in that the latter is the unique irreducible quotient of the former.  We have also shown that the affine mode $h_0$ is diagonalisable on both $\AffStagMod{0}$ and $\AffStagMod{1}$, but $L_0$ is not.  Indeed, the non-diagonalisable action of $L_0$ links the states of the socles with their Jordan partners, the latter being associated with the composition factors in the top row of the above diagram.

Because of this structure, the $\AffStagMod{\lambda}$ may also be described as \emph{staggered} modules in the spirit of \cite{RohRed96,RidSta09}.  Combining the composition factors in the diagrams above along the south-east arrows, we obtain exact sequences
\begin{subequations}
\begin{align}
&\dses{\tfunc{\gamma^{-1}}{\AffOthMod{-1/2}^+}}{\AffStagMod{0}}{\tfunc{\gamma}{\AffOthMod{-3/2}^+}} \label{SESS0} \\
\text{and} \qquad &\dses{\tfunc{\gamma^{-1}}{\AffOthMod{-3/2}^+}}{\AffStagMod{1}}{\tfunc{\gamma}{\AffOthMod{-1/2}^+}}.
\end{align}
\end{subequations}
One can obtain similar exact sequences involving the $\AffOthMod{\mu}^-$ by combining the composition factors along the south-west arrows.  We have also demonstrated that the structures described here completely determine the $\AKMA{sl}{2}$-action on the $\AffStagMod{\lambda}$ (see \secDref{secAnalysis1}{secAnalysis2} for the explicit formulae).  This is of some interest because it was claimed in \cite{GabFus01} that this was not the case for at least one of the indecomposables encountered in the $k = -\tfrac{4}{3}$ fractional level model.  More precisely, the statement there is that the structure of this module was fixed up to an unknown constant, with different constants parametrising non-isomorphic modules.  It would be very interesting to understand if there is a structural reason behind this difference between the $k = -\tfrac{1}{2}$ and $k = -\tfrac{4}{3}$ cases, similar to that observed in \cite{RidLog07,RidSta09} for the Virasoro algebra.  In any case, it is germane to ask if there is a theory of staggered modules for $\AKMA{sl}{2}$ analogous to the Virasoro story.

For completeness, it is worth mentioning that we have illustrated the structure of the indecomposables $\AffStagMod{0}$ and $\AffStagMod{1}$ in \figDref{figS0}{figS1}.  These pictures may be directly compared to the ``extremal diagrams'' of the indecomposables constructed in \cite{LesLog04} from a free field construction.  It appears that we have found agreement, although their version of $\AffStagMod{1}$ is only half complete and their diagrams seem to attach an undue importance to the states of conformal dimension $0$ and $\tfrac{1}{2}$ (most of which are in no way extremal).  One may therefore view the results reported here as a clarification and confirmation of their results.  In particular, our description of the indecomposable structure refines the character formulae given in \cite{LesLog04}.

We mention some further observations that may be of interest.  First, the $\AffOthMod{\lambda}$, the $\AffStagMod{\lambda}$ and their twisted versions form an ideal of the fusion ring, suggesting that they may be \emph{projective} in the category of admissible $k = -\tfrac{1}{2}$ $\AKMA{sl}{2}$-modules.  If true, this would give a simple proof that the decomposition of the fusion rule \eqref{eqnFusES0} is direct.  We note that quotienting the fusion ring by this ideal results in the fusion ring of the non-logarithmic theory discussed in \cite{LesSU202,RidSL208}.

Second, one has come to expect that the fusion of staggered modules may be computed by temporarily forgetting some of the indecomposable structure, computing some more simple fusions, and then reconstituting appropriate indecomposable structures in the results.\footnote{This expectation arises in the consideration of whether the fusion product descends to the Grothendieck ring of characters.  However, it is more fundamental than the character product when the kernel of the map from modules to characters is large.}  In particular, the exact sequence \eqref{SESS0} for $\AffStagMod{0}$ suggests that
\begin{equation}
\AffOthMod{\lambda} \fuse \AffStagMod{0} = \AffOthMod{\lambda} \fuse \tfunc{\gamma^{-1}}{\AffOthMod{-1/2}^+} + \AffOthMod{\lambda} \fuse \tfunc{\gamma}{\AffOthMod{-3/2}^+},
\end{equation}
where the ``$+$'' indicates that we may be forgetting some indecomposable structure.  Assuming that \eqnref{eqnFusEEFull} extends to $\mu \in \ZZ + \tfrac{1}{2}$ (perhaps with some additional indecomposable structure), this suggests that $\AffOthMod{\lambda} \fuse \AffStagMod{0}$ should decompose into the \emph{four} irreducibles $\AffOthMod{\lambda}$, $\tfunc{\gamma^{-2}}{\AffOthMod{\lambda - 1}}$, $\tfunc{\gamma^2}{\AffOthMod{\lambda - 1}}$ and $\AffOthMod{\lambda}$ (at least at the level of composition factors), rather than just two as \eqnref{eqnFusES0'} originally concluded.  In fact, this expectation predicts the fusion result \eqref{eqnFusES0}.  Indeed, it was this which originally prompted the additional computation to grade $\brac{0,0,2}$ in \secref{secFusES}.

Third, the fusion ring \eqref{eqnFusion} shows significant similarities to the fusion ring of the $c=-2$ triplet model as given in \cite{GabRat96}.  For completeness, we note that this ring is generated by four irreducibles denoted by $\mathcal{V}_0$, $\mathcal{V}_1$, $\mathcal{V}_{-1/8}$ and $\mathcal{V}_{3/8}$.  There are, in addition, two indecomposables which are denoted by $\mathcal{R}_0$ and $\mathcal{R}_1$.  The fusion rules are as follows:  $\mathcal{V}_0$ is the fusion identity and
\begin{align} \label{eqnTriplet}
&
\begin{aligned}
\mathcal{V}_1 \fuse \mathcal{V}_1 &= \mathcal{V}_0, \\
\mathcal{V}_1 \fuse \mathcal{V}_{-1/8} &= \mathcal{V}_{3/8}, \\
\mathcal{V}_1 \fuse \mathcal{V}_{3/8} &= \mathcal{V}_{-1/8}, \\
\mathcal{V}_1 \fuse \mathcal{R}_0 &= \mathcal{R}_1, \\
\mathcal{V}_1 \fuse \mathcal{R}_1 &= \mathcal{R}_0,
\end{aligned}
& &
\begin{aligned}
\mathcal{V}_{-1/8} \fuse \mathcal{V}_{-1/8} &= \mathcal{R}_0, \\
\mathcal{V}_{-1/8} \fuse \mathcal{V}_{3/8} &= \mathcal{R}_1, \\
\mathcal{V}_{3/8} \fuse \mathcal{V}_{3/8} &= \mathcal{R}_0, \\
\mathcal{V}_h \fuse \mathcal{R}_{\lambda} &= 2 \mathcal{V}_{-1/8} \oplus 2 \mathcal{V}_{3/8} & \text{($h = -\tfrac{1}{8} , \tfrac{3}{8}$; $\lambda = 0,1$),} \\
\mathcal{R}_{\lambda} \fuse \mathcal{R}_{\mu} &= 2 \mathcal{R}_0 \oplus 2 \mathcal{R}_1 & \text{($\lambda = 0,1$).}
\end{aligned}
\end{align}
The relation between the fusion rules \eqref{eqnFusion} and \eqref{eqnTriplet} amounts to merely neglecting the spectral flow.  More precisely, if we let $\sqbrac{\mathcal{M}}$ denote the equivalence class of all spectral flow images of the $\AKMA{sl}{2}$-module $\mathcal{M}$, then the relation becomes a ring isomorphism given by
\begin{subequations}
\begin{align}
\bigl[ \AffIrrMod{0} \bigr] &\longleftrightarrow \mathcal{V}_0, & 
\bigl[ \AffOthMod{0} \bigr] &\longleftrightarrow \mathcal{V}_{-1/8}, &
\bigl[ \AffStagMod{0} \bigr] &\longleftrightarrow \mathcal{R}_0, \\
\bigl[ \AffIrrMod{1} \bigr] &\longleftrightarrow \mathcal{V}_1, &
\bigl[ \AffOthMod{1} \bigr] &\longleftrightarrow \mathcal{V}_{3/8}, &
\bigl[ \AffStagMod{1} \bigr] &\longleftrightarrow \mathcal{R}_1.
\end{align}
\end{subequations}
This isomorphism gives us confidence that the $\AKMA{sl}{2}_{-1/2}$ fusion rules reported here (and the $c=-2$ triplet model fusion rules reported in \cite{GabRat96}) are correct.  Of course, we should expect such a relation to hold, given the realisation of the triplet model as the $\AKMA{u}{1}$-coset of the (logarithmic) $\AKMA{sl}{2}_{-1/2}$ theory \cite{RidSL210}.  However, the familiar argument from rational \cft{} which would guarantee the above relation --- computing the fusion rules of a coset theory from the modular properties of its characters and the Verlinde formula --- \emph{does not apply}, because the fusion ring cannot, in this case, be reconstructed from the modular transformations.  This relation therefore requires a more fundamental (and probably more natural) explanation.

We conclude by briefly discussing the implications of these results for the $\beta \gamma$ ghost system \eqref{eqnGhostComm}.  As mentioned in \secref{secOld}, this algebra corresponds to an extension of the affine Kac-Moody algebra by the (zero-grade fields of the) simple current $\AffIrrMod{1}$.  The orbits in the fusion ring \eqref{eqnFusion} under the simple current action therefore combine into modules for the $\beta \gamma$ system.  Specifically, we find two families of irreducible $\beta \gamma$-modules $\tfunc{\gamma^{\ell}}{\mathsf{L}}$ and $\tfunc{\gamma^{\ell}}{\mathsf{E}}$, and a single family of indecomposables $\tfunc{\gamma^{\ell}}{\mathsf{S}}$, confirming the logarithmic nature of the $\beta \gamma$ ghost system.  We picture the $\ell = 0$ representatives of these families in \figref{figGhostMods}.  Their structure and fusion rules are easily deduced from the results presented here.  For example, $\mathsf{L}$ is found to be the fusion identity, whereas
\begin{equation}
\mathsf{E} \fuse \mathsf{E} = \mathsf{S}, \qquad \mathsf{E} \fuse \mathsf{S} = \tfunc{\gamma^{-2}}{\mathsf{E}} \oplus 2 \mathsf{E} \oplus \tfunc{\gamma^2}{\mathsf{E}} \qquad \text{and} \qquad \mathsf{S} \fuse \mathsf{S} = \tfunc{\gamma^{-2}}{\mathsf{S}} \oplus 2 \mathsf{S} \oplus \tfunc{\gamma^2}{\mathsf{S}}.
\end{equation}
Note however that the $\beta \gamma$ chiral algebra admits a much larger spectrum, so that which appears here (and in \cite{RidSL208}) must correspond to a compactification/orbifold.  We hope to return to this in the future.

{
\psfrag{L0}[][]{$\mathsf{L}$}
\psfrag{E0}[][]{$\mathsf{E}$}
\psfrag{S0}[][]{$\mathsf{S}$}
\psfrag{00}[][]{$\scriptstyle \brac{0,0}$}
\psfrag{0e}[][]{$\scriptstyle \brac{0,-\tfrac{1}{8}}$}
\psfrag{aa}[][]{$\scriptstyle \brac{\tfrac{1}{2},-\tfrac{1}{8}}$}
\psfrag{ee}[][]{$\scriptstyle \brac{1,\tfrac{1}{2}}$}
\psfrag{ff}[][]{$\scriptstyle \brac{-1,\tfrac{1}{2}}$}
\psfrag{eE}[][]{$\scriptstyle \brac{1,-\tfrac{1}{8}}$}
\psfrag{fE}[][]{$\scriptstyle \brac{-1,-\tfrac{1}{8}}$}
\psfrag{a}[][]{$\scriptstyle \brac{1,-\tfrac{1}{2}}$}
\psfrag{b}[][]{$\scriptstyle \brac{-1,-\tfrac{1}{2}}$}
\begin{figure}
\begin{center}
\includegraphics[width=0.9\textwidth]{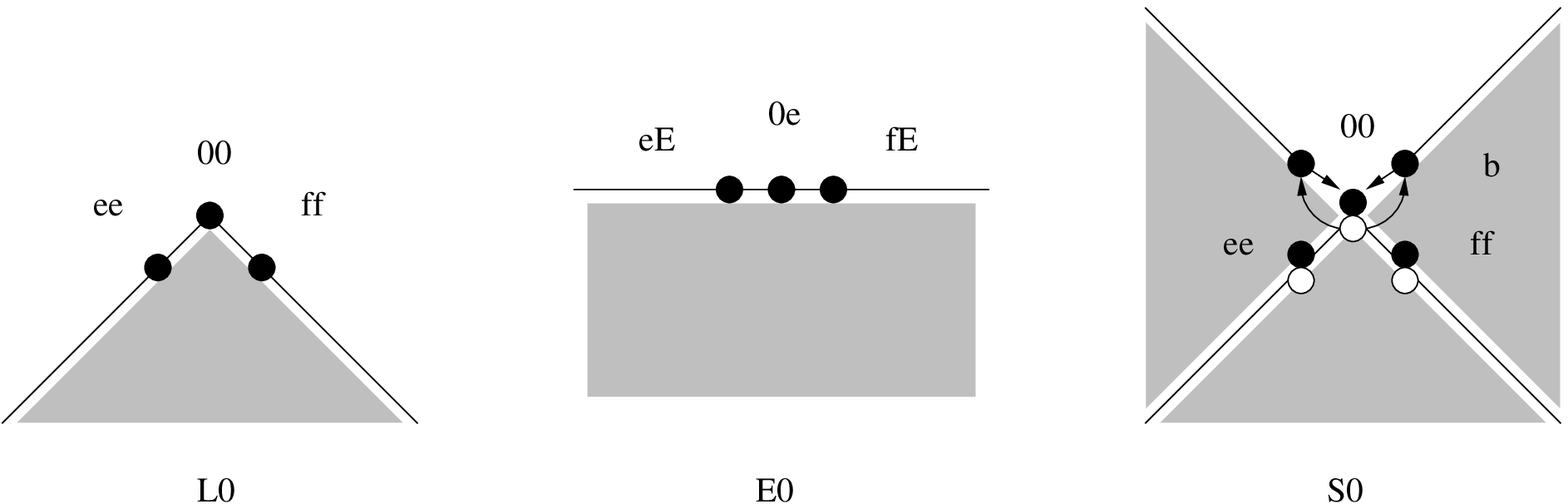}
\caption{Depictions of the untwisted $\beta \gamma$-modules obtained from the spectra of our $\AKMA{sl}{2}_{-1/2}$-theories.  Each labelled state declares its $\func{\alg{sl}}{2}$-weight and conformal dimension (in that order).  The ghosts themselves are the dimension $\tfrac{1}{2}$ fields appearing in $\mathsf{L} \subset \mathsf{S}$.} \label{figGhostMods}
\end{center}
\end{figure}
}

\section*{Acknowledgements}

I would like to thank Matthias Gaberdiel for originally pointing out that twisted representations can lurk, unseen, in the shadows of zero-grade fusion computations.  The results presented here have also benefitted enormously from discussions with Thomas Creutzig, Pierre Mathieu and Yvan Saint-Aubin.  This research was supported under the Australian Research Council's Discovery Projects funding scheme (project number DP1093910).


\begin{thebibliography}{10}

\bibitem{RidSL208}
D~Ridout.
\newblock {$\widehat{\mathfrak{sl}} \left( 2 \right)_{-1/2}$: A Case Study}.
\newblock {\em Nucl. Phys.}, B814:485--521, 2009.
\newblock \texttt{arXiv:0810.3532 [hep-th]}.

\bibitem{RidSL210}
D~Ridout.
\newblock {$\widehat{\mathfrak{sl}} \left( 2 \right)_{-1/2}$ and the Triplet
  Model}.
\newblock {\em Nucl. Phys.}, B835:314--342, 2010.
\newblock \texttt{arXiv:1001.3960 [hep-th]}.

\bibitem{RozQua92}
L~Rozansky and H~Saleur.
\newblock {Quantum Field Theory for the Multivariable Alexander-Conway
  Polynomial}.
\newblock {\em Nucl. Phys.}, B376:461--509, 1992.

\bibitem{GurLog93}
V~Gurarie.
\newblock {Logarithmic Operators in Conformal Field Theory}.
\newblock {\em Nucl. Phys.}, B410:535--549, 1993.
\newblock \texttt{arXiv:hep-th/9303160}.

\bibitem{LesSU202}
F~Lesage, P~Mathieu, J~Rasmussen, and H~Saleur.
\newblock {The $\widehat{su} \left( 2 \right)_{-1/2}$ WZW Model and the $\beta
  \gamma$ System}.
\newblock {\em Nucl. Phys.}, B647:363--403, 2002.
\newblock \texttt{arXiv:hep-th/0207201}.

\bibitem{LesLog04}
F~Lesage, P~Mathieu, J~Rasmussen, and H~Saleur.
\newblock {Logarithmic Lift of the $\widehat{su} \left( 2 \right)_{-1/2}$
  Model}.
\newblock {\em Nucl. Phys.}, B686:313--346, 2004.
\newblock \texttt{arXiv:hep-th/0311039}.

\bibitem{NahQua94}
W~Nahm.
\newblock {Quasirational Fusion Products}.
\newblock {\em Int. J. Mod. Phys.}, B8:3693--3702, 1994.
\newblock \texttt{arXiv:hep-th/9402039}.

\bibitem{GabInd96}
M~Gaberdiel and H~Kausch.
\newblock {Indecomposable Fusion Products}.
\newblock {\em Nucl. Phys.}, B477:293--318, 1996.
\newblock \texttt{arXiv:hep-th/9604026}.

\bibitem{GabInt00}
M~Gaberdiel.
\newblock {An Introduction to Conformal Field Theory}.
\newblock {\em Rep. Prog. Phys.}, 63:607--667, 2000.
\newblock \texttt{arXiv:hep-th/9910156}.

\bibitem{GabAlg03}
M~Gaberdiel.
\newblock {An Algebraic Approach to Logarithmic Conformal Field Theory}.
\newblock {\em Int. J. Mod. Phys.}, A18:4593--4638, 2003.
\newblock \texttt{arXiv:hep-th/0111260}.

\bibitem{EbeVir06}
H~Eberle and M~Flohr.
\newblock {Virasoro Representations and Fusion for General Augmented Minimal
  Models}.
\newblock {\em J. Phys.}, A39:15245--15286, 2006.
\newblock \texttt{arXiv:hep-th/0604097}.

\bibitem{RidPer07}
P~Mathieu and D~Ridout.
\newblock {From Percolation to Logarithmic Conformal Field Theory}.
\newblock {\em Phys. Lett.}, B657:120--129, 2007.
\newblock \texttt{arXiv:0708.0802 [hep-th]}.

\bibitem{RidLog07}
P~Mathieu and D~Ridout.
\newblock {Logarithmic $M \left( 2,p \right)$ Minimal Models, their Logarithmic
  Couplings, and Duality}.
\newblock {\em Nucl. Phys.}, B801:268--295, 2008.
\newblock \texttt{arXiv:0711.3541 [hep-th]}.

\bibitem{RidPer08}
D~Ridout.
\newblock {On the Percolation BCFT and the Crossing Probability of Watts}.
\newblock {\em Nucl. Phys.}, B810:503--526, 2009.
\newblock \texttt{arXiv:0808.3530 [hep-th]}.

\bibitem{RidSta09}
K~Kyt\"{o}l\"{a} and D~Ridout.
\newblock {On Staggered Indecomposable Virasoro Modules}.
\newblock {\em J. Math. Phys.}, 50:123503, 2009.
\newblock \texttt{arXiv:0905.0108 [math-ph]}.

\bibitem{GabFus09}
M~Gaberdiel, I~Runkel, and S~Wood.
\newblock {Fusion Rules and Boundary Conditions in the $c=0$ Triplet Model}.
\newblock {\em J. Phys.}, A42:325403, 2009.
\newblock \texttt{arXiv:0905.0916 [hep-th]}.

\bibitem{GabFus97}
M~Gaberdiel.
\newblock {Fusion of Twisted Representations}.
\newblock {\em Int. J. Mod. Phys.}, A12:5183--5208, 1997.
\newblock \texttt{arXiv:hep-th/9607036}.

\bibitem{GabFus01}
M~Gaberdiel.
\newblock {Fusion Rules and Logarithmic Representations of a WZW Model at
  Fractional Level}.
\newblock {\em Nucl. Phys.}, B618:407--436, 2001.
\newblock \texttt{arXiv:hep-th/0105046}.

\bibitem{GabRat96}
M~Gaberdiel and H~Kausch.
\newblock {A Rational Logarithmic Conformal Field Theory}.
\newblock {\em Phys. Lett.}, B386:131--137, 1996.
\newblock \texttt{arXiv:hep-th/9606050}.

\bibitem{WooFus10}
S~Wood.
\newblock {Fusion Rules of the $W \left( p,q \right)$ Triplet Models}.
\newblock {\em J. Phys.}, A43:045212, 2010.
\newblock \texttt{arXiv:0907.4421 [hep-th]}.

\bibitem{FucNon04}
J~Fuchs, S~Hwang, A~Semikhatov, and I~Yu Tipunin.
\newblock {Nonsemisimple Fusion Algebras and the Verlinde Formula}.
\newblock {\em Comm. Math. Phys.}, 247:713--742, 2004.
\newblock \texttt{arXiv:hep-th/0306274}.

\bibitem{PeaLog06}
P~Pearce, J~Rasmussen, and J-B~Zuber.
\newblock {Logarithmic Minimal Models}.
\newblock {\em J. Stat. Mech.}, 0611:017, 2006.
\newblock \texttt{arXiv:hep-th/0607232}.

\bibitem{ReaAss07}
N~Read and H~Saleur.
\newblock {Associative-Algebraic Approach to Logarithmic Conformal Field
  Theories}.
\newblock {\em Nucl. Phys.}, B777:316--351, 2007.
\newblock \texttt{arXiv:hep-th/0701117}.

\bibitem{HuaLog07}
Y-Z Huang, J~Lepowsky, and L~Zhang.
\newblock {Logarithmic Tensor Product Theory for Generalized Modules for a
  Conformal Vertex Algebra}.
\newblock \texttt{arXiv:0710.2687 [math.QA]}.

\bibitem{PeaInt08}
P~Pearce, J~Rasmussen, and P~Ruelle.
\newblock {Integrable Boundary Conditions and $W$-Extended Fusion in the
  Logarithmic Minimal Models $LM \left( 1,p \right)$}.
\newblock {\em J. Phys.}, A41:295201, 2008.
\newblock \texttt{arXiv:0803.0785 [hep-th]}.

\bibitem{KytFro08}
K~Kyt\"{o}l\"{a}.
\newblock {From SLE to the Operator Content of Percolation}.
\newblock {\em J. Stat. Mech.}, 0908:P08005, 2009.
\newblock \texttt{arXiv:0804.2612 [math-ph]}.

\bibitem{RohRed96}
F~Rohsiepe.
\newblock {On Reducible but Indecomposable Representations of the Virasoro
  Algebra}.
\newblock \texttt{arXiv:hep-th/9611160}.

\bibitem{KacMod88b}
V~Kac and M~Wakimoto.
\newblock {Modular and Conformal Invariance Constraints in Representation
  Theory of Affine Algebras}.
\newblock {\em Adv. Math.}, 70:156--236, 1988.

\bibitem{FeiEqu98}
B~Feigin, A~Semikhatov, and I~Yu Tipunin.
\newblock {Equivalence Between Chain Categories of Representations of Affine
  $sl \left( 2 \right)$ and $N = 2$ Superconformal Algebras}.
\newblock {\em J. Math. Phys.}, 39:3865--3905, 1998.
\newblock \texttt{arXiv:hep-th/9701043}.

\bibitem{SemEmb97}
A~Semikhatov and V~Sirota.
\newblock {Embedding Diagrams of $N=2$ Verma Modules and Relaxed $\widehat{sl}
  \left( 2 \right)$ Verma Modules}.
\newblock \texttt{arXiv:hep-th/9712102}.

\bibitem{DiFCon97}
P~Di Francesco, P~Mathieu, and D~S\'{e}n\'{e}chal.
\newblock {\em {Conformal Field Theory}}.
\newblock Graduate Texts in Contemporary Physics. Springer-Verlag, New York,
  1997.

\bibitem{FeiAnn92}
B~Feigin, T~Nakanishi, and H~Ooguri.
\newblock {The Annihilating Ideals of Minimal Models}.
\newblock {\em Int. J. Mod. Phys.}, A7:217--238, 1992.

\bibitem{GabFus94}
M~Gaberdiel.
\newblock {Fusion in Conformal Field Theory as the Tensor Product of the
  Symmetry Algebra}.
\newblock {\em Int. J. Mod. Phys.}, A9:4619--4636, 1994.
\newblock \texttt{arXiv:hep-th/9307183}.

\end{thebibliography}
\end{document}